\documentclass[sigconf]{acmart}

\AtBeginDocument{%
  }

\copyrightyear{2023}
\acmYear{2023}
\setcopyright{rightsretained}
\acmConference[MICRO '23]{56th Annual IEEE/ACM International Symposium on Microarchitecture}{October 28-November 1, 2023}{Toronto, ON, Canada}
\acmBooktitle{56th Annual IEEE/ACM International Symposium on Microarchitecture (MICRO '23), October 28-November 1, 2023, Toronto, ON, Canada}
\acmDOI{10.1145/3613424.3623793}
\acmISBN{979-8-4007-0329-4/23/10}
\usepackage{multirow, booktabs}
\usepackage{amsmath}
\usepackage{caption}
\usepackage{subcaption}
\usepackage{enumitem}
\usepackage{tikz}
\usepackage{yfonts}
\usepackage{dutchcal}
\usepackage{wrapfig}
\usepackage{floatflt}
\usepackage{enumitem}
\usepackage{float}
\usepackage{algorithm}
\usepackage{subcaption}
\usepackage{algpseudocode}
\setlist{nosep}
\newcommand\blfootnote[1]{%
  \begingroup
  \renewcommand\thefootnote{}\footnote{#1}%
  \addtocounter{footnote}{-1}%
  \endgroup
}
\definecolor{BrickRed}{HTML}{B6321C}
\definecolor{Green}{HTML}{00A64F}
\definecolor{LimeGreen}{HTML}{8DC73E}

\newcommand{\ie}{{\em i.e.}}
\newcommand{\eg}{{\em e.g.}}
\newcommand{\obsb}{Tailors}
\newcommand{\tilingName}{Swiftiles}  

\begin{document}

\title{Tailors: Accelerating Sparse Tensor Algebra by Overbooking Buffer Capacity} 

\author{Zi Yu Xue}
\affiliation{%
  \institution{MIT}
  \city{Cambridge}
  \state{MA}
  \country{USA}
}
\email{fzyxue@mit.edu}

\author{Yannan Nellie Wu}
\affiliation{%
  \institution{MIT}
  \city{Cambridge}
  \state{MA}
  \country{USA}
}
\email{nelliewu@mit.edu}

\author{Joel S. Emer}
\affiliation{%
  \institution{MIT/NVIDIA}
  \city{Cambridge}
  \state{MA}
  \country{USA}
}
\email{jsemer@mit.edu}

\author{Vivienne Sze}
\affiliation{%
  \institution{MIT}
  \city{Cambridge}
  \state{MA}
  \country{USA}
}
\email{sze@mit.edu}

\begin{abstract}
Sparse tensor algebra is a challenging class of workloads to accelerate due to low arithmetic intensity and varying sparsity patterns. Prior sparse tensor algebra accelerators have explored tiling sparse data to increase exploitable data reuse and improve throughput, but typically allocate tile size in a given buffer for the worst-case data occupancy. This severely limits the utilization of available memory resources and reduces data reuse. Other accelerators employ complex tiling during preprocessing or at runtime to determine the exact tile size based on its occupancy. 

This paper proposes a speculative tensor tiling approach, called \emph{overbooking}, to improve buffer utilization by taking advantage of the distribution of nonzero elements in sparse tensors to construct larger tiles with greater data reuse. To ensure correctness, we propose a low-overhead hardware mechanism, \emph{\obsb}, that can tolerate data overflow by design while ensuring reasonable data reuse. We demonstrate that \obsb \ can be easily integrated into the memory hierarchy of an existing sparse tensor algebra accelerator. To ensure high buffer utilization with minimal tiling overhead, we introduce a statistical approach, \emph{\tilingName}, to pick a tile size so that tiles usually fit within the buffer's capacity, but can potentially overflow, \ie, it \emph{overbooks} the buffers. Across a suite of 22 sparse tensor algebra workloads, we show that our proposed overbooking strategy introduces an average speedup of $52.7\times$ and $2.3\times$ and an average energy reduction of $22.5\times$ and $2.5\times$ over ExTensor without and with optimized tiling, respectively.
\end{abstract}

\maketitle



\section{Introduction}

\label{sec:intro}

\renewcommand{\arraystretch}{1.4}
\begin{table}[tb]
\centering
\resizebox{\columnwidth}{!}{
\begin{tabular}{c|c|c}

        \textbf{ Tiling Strategies}            & \textbf{Buffer Utilization} & \textbf{Tiling Tax} \\ \hline
\multicolumn{1}{c|}{Uniform shape}    & {\color{BrickRed}{\textbf{Very Low}}}                & {\color{Green}{\textbf{None}}}                 \\ \hline

\multicolumn{1}{c|}{Prescient uniform shape}      & \color{BrickRed}{\textbf{Low}}         & \color{BrickRed}{\textbf{High}}                 \\ \hline

\multicolumn{1}{c|}{Uniform occupancy}       & \color{LimeGreen}{\textbf{High}}                    & \color{BrickRed}{\textbf{Very High}}                 \\ \hline

\multicolumn{1}{c|}{\textbf{Our work (Overbooking)}}            & \color{LimeGreen}{\textbf{High}}               & \color{LimeGreen}{\textbf{Low}}                   \\ \hline
\end{tabular}
}
\caption{Comparison of different tiling strategies. For any workload, an ideal tiling strategy should achieve high buffer utilization and introduce low tiling tax.}

\label{tab:existing-work}
\vspace{-16pt}

\end{table}
\renewcommand{\arraystretch}{1.0}

Tensor algebra is a computing paradigm that is used across a wide variety of application domains, \eg, graph computing~\cite{sparse-graph, social-networks-graph}, scientific simulations~\cite{power-simulation, fluiddynamics}, data analytics ~\cite{elgendy_bigdata_2014}, and recommendation systems~\cite{rivera_recommendation_2018}. Many of these applications operate on large tensors that have high sparsity. In addition, the distributions of zero-value locations vary significantly both across tensors and within a single tensor.  For example, a road network graph like \textit{roadNet-CA}~\cite{kolodziej_suitesparse_2019} has an adjacency matrix potentially containing trillions of elements but with only millions of them being nonzeros. These nonzeros are densely populated along the diagonal and randomly scattered away from the diagonal. This large tensor size and high sparsity limit the arithmetic intensity (\ie, the ratio of operations to data traffic from DRAM) and make the processing of these large sparse tensors challenging and memory bound on most architectures. \blfootnote{More information on Tailors can be found at \url{http://emze.csail.mit.edu/tailors}.}

A popular approach to address the above challenge is to partition large tensors into smaller sub-tensors, referred to as \emph{tiles}, which can be stored in smaller buffers for reuse to significantly reduce memory traffic~\cite{hegde_extensor_2019}. Tiles can have: \textbf{i)} different \emph{shapes} described by a tuple of ranges (\ie, the number of elements including both zeros and nonzeros, along each dimension). For example, a two-dimensional tile can have a 4-by-4 shape, which has a range of four along each dimension); \textbf{ii)} different \emph{sizes} (\ie, the number of elements including both zeros and nonzeros, for example, a 4-by-4 tile has a size of 16); \textbf{iii)} different \emph{occupancies} (\ie, number of nonzero elements). For sparse tensors, an ideal tiling would use the \emph{largest tile size} for which the data fits within a buffer to maximize buffer utilization (\ie, the percentage of the buffer occupied by data) and thus data reuse, while simultaneously constructing the \emph{smallest buffer} to minimize energy and area. Specifically, an ideal tiling strategy has the following goals:
\begin{itemize}
    \item \textbf{Adaptability}: always achieves high buffer utilization across a wide range of different sparsity distributions (\ie, spatial distribution of nonzero values and degrees of sparsity).

    \item \textbf{Efficiency}: partitions the tensors into tiles with low pre-processing or runtime cost and low hardware cost for \emph{operand matching} (\ie, when given a tile of an operand tensor $A$, find the corresponding range of coordinates in the other operand tensor $B$). We refer to such overhead as the \emph{tiling tax}. 

\end{itemize}

However, to the authors' knowledge, no tiling strategy employed by existing sparse tensor accelerators achieves both goals. Existing schemes generally select tile sizes by statically selecting a shape that is known not to overflow the buffers~\cite{hegde_extensor_2019, matraptor, sigma, outerspace, sparch} or filling buffers with a runtime-determined shape that will fit~\cite{odemuyiwa_drt_2023, GAMMA-accelerator}. As introduced in~\cite{odemuyiwa_drt_2023}, two common strategies for tiling tensors are tiling with \emph{uniform shape} tiles and tiling with \emph{uniform occupancy} tiles. We summarize the characteristics of different tiling strategies in Table~\ref{tab:existing-work}.

\begin{figure}[t]%
    \centering
    \hspace{-10pt}
    \vspace{-10pt}

    \includegraphics[width=1.08\columnwidth]{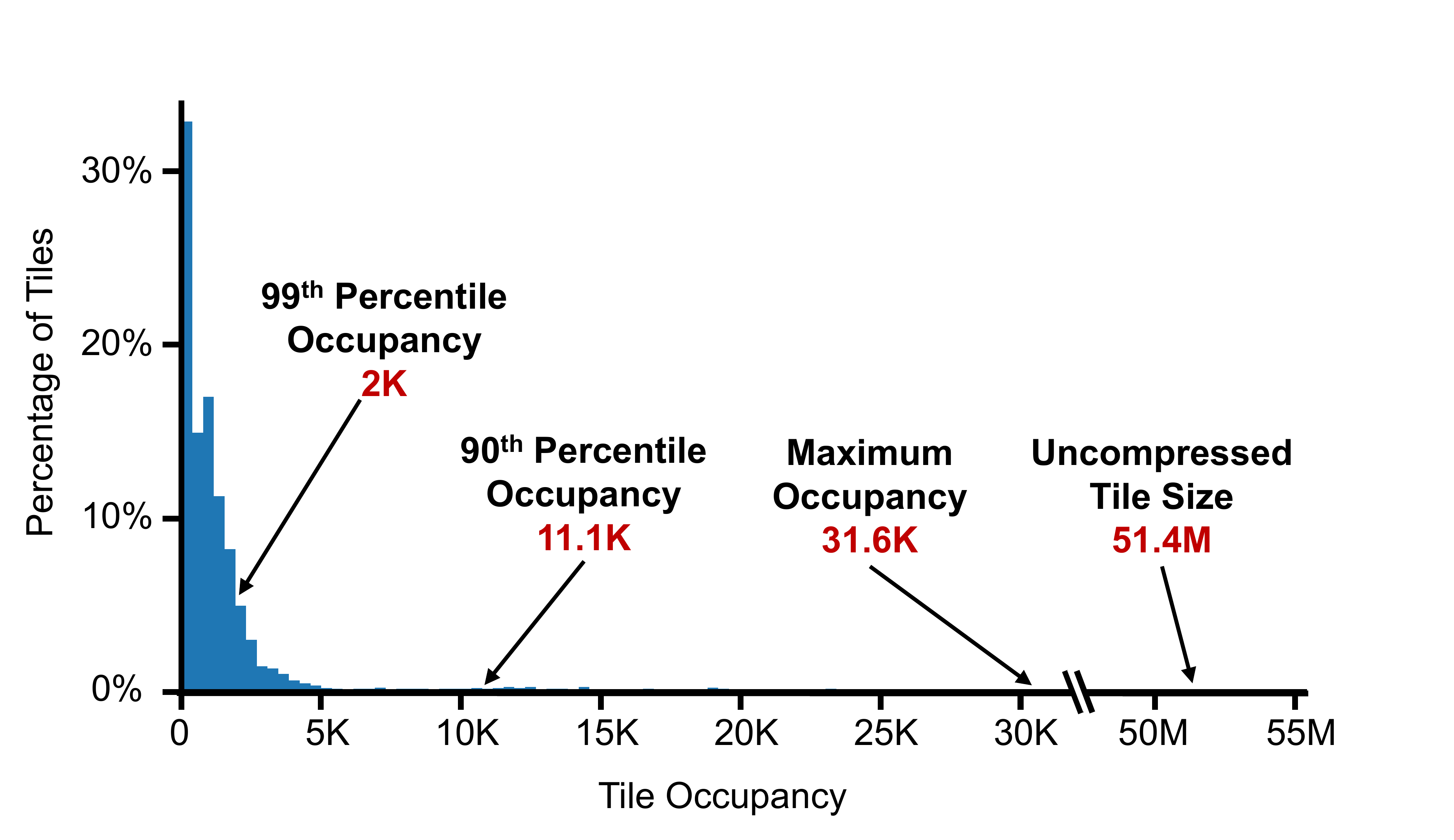} 
    \vspace{-20pt}
    
    \caption{\textbf{Occupancy distribution of tiles with a size of 51.4M. The tiles are obtained by partitioning tensors from SuiteSparse~\cite{kolodziej_suitesparse_2019}. The occupancy varies from tile to tile, the max tile occupancy is more than three orders of magnitude smaller than tile size, and 90\% threshold tile occupancy is more than $15\times$ smaller than maximum tile occupancy.}}
    \label{fig:longtail}
\end{figure}

The \emph{uniform shape} tiling strategy, which statically selects a shape to tile with, partitions a tensor into tiles of identical shapes based on the available buffer capacity without regard to tensor sparsity.  In particular, not being aware of tensor sparsity, the uniform shape tiling assumes worst-case occupancy (\ie, assumes a dense tile) and thus mandates the tile size to not exceed the available buffer capacity. Since the tile shapes are fixed, uniform shape tiling does not require any runtime overhead for operand matching, thus introducing zero tiling tax. However, since sparse tensor algebra workloads often have high sparsity, uniform shape tiling's extremely conservative strategy can often result in severely underutilized buffers. For example, Fig.~\ref{fig:longtail} shows the tile occupancy distribution when a tensor from the SuiteSparse dataset~\cite{kolodziej_suitesparse_2019} is tiled using the uniform shape strategy. Although the worst-case occupancy is 51.4M elements, the maximum tile occupancy observed in the tensor is only 31.6K, thus resulting in, at best, a less than $0.1\%$ average buffer utilization.

Uniform shape tiling can be enhanced to take tensor sparsity into account instead of constructing tiles based on the worst-case occupancy. We refer to this enhanced tiling strategy as \emph{prescient uniform shape} tiling, which partitions the tensor based on prescient knowledge about the maximum tile occupancy. Specifically, prescient uniform-shape tiling partitions the tensor into tiles with a larger size as long as the maximum occupancy among such tiles does not exceed the available buffer capacity. However, such an approach introduces significant pre-processing cost when the tensor is static, or a runtime cost when the tensor is generated during execution, thus has a high tiling tax (\eg, for each workload, all tile shapes of interest need to be explored, and for each tile shape, the maximum tile occupancy needs to be measured, which requires traversing the entire tensor). In addition, we observe that even with prescient uniform shape tiling, the buffer utilization is still low as the tile occupancy varies significantly from tile to tile and the maximum tile occupancy is often much larger than that of the majority of the tiles.  For example, Fig.~\ref{fig:longtail} shows that while the maximum occupancy among the tiles is 31.6K, 90\% of the tiles have occupancies of less than 2K; this leads to a buffer utilization that is less than 10\% for 90\% of the time. Thus, as shown in Table~\ref{tab:existing-work}, prescient uniform shape introduces an undesirable tradeoff between tiling tax and buffer utilization.

As an alternative to the uniform shape approach, the \emph{uniform occupancy} tiling strategy aims to improve buffer utilization by constructing tiles based on the exact number of nonzero values in the tensor. Ideally, uniform occupancy tiling aims to always fully utilize the available buffer capacity with tiles that have the perfect number of nonzeros to fill every buffer. However, uniform occupancy tiling often results in non-uniform shapes among the tiles, especially when the nonzero value distribution is not uniform, leading to significant tiling tax associated with runtime operand matching. In addition, due to the complexity involved in operand matching, existing work can only emulate uniform occupancy tiling behaviors with tiles that have occupancies that are similar, but smaller than, the available buffer capacity, and thus cannot achieve perfect buffer utilization~\cite{odemuyiwa_drt_2023}.

To address the above limitations, we propose a simultaneously adaptable and efficient tiling strategy, with the key insight that workload tensors can be partitioned into uniformly shaped tiles that \emph{sometimes require a larger buffer capacity than is available}. We refer to such tiling strategy as \emph{overbooking}.\footnote{We choose to name our strategy overbooking due to similarities with how airlines sell more tickets (larger tile shape) for a flight than the plane (buffer) has capacity for, thus potentially "overbooking" the plane with more passengers (nonzeros) than it can hold.} In particular, the overbooking tiling strategy speculatively constructs uniformly shaped tiles, such that approximately $y\%$ of the tiles will \textbf{not} fit into the buffer, referred to as a $y\%$ overbooking. As shown in Table~\ref{tab:existing-work}, our proposed overbooking strategy is both adaptable and efficient. At a high level, our proposal achieves both goals by: \textbf{i)} implementing low-cost hardware support called \emph{\obsb} that turns data reuse into data streaming to guarantee correctness and throughput while maintaining some data reuse when a buffer is overbooked; and \textbf{ii)} employing an overbooking tiling strategy called \emph{\tilingName}, which employs low-overhead statistical characterizations of the tensor sparsity to pick a tile size that leads to high buffer utilization, but can be overbooked $y\%$ of the time.

Table~\ref{tab:existing-work} summarizes the different tiling strategies in terms of their adaptability (measured by buffer utilization), and efficiency (measured by tiling tax).

This work makes the following key contributions:
\begin{itemize}
\item This is the first work to demonstrate that the concept of speculative execution can be applied to tiling sparse tensors by \emph{overbooking} buffer capacities.

\item To ensure correctness for an overbooked buffer, we propose \emph{\obsb}, a low-cost hardware mechanism that streams the overflowed data with a low-cost latency hiding queue while maintaining reasonable data reuse. 
\item We show that \obsb \ can be easily integrated into the memory hierarchy of an existing accelerator.
\item To balance tiling efficiency and adaptability, we propose \emph{Swiftiles}, which swiftly determines the tile size of the sparse tensors statistically by sampling the irregular distribution of real-world data.
\item Across a suite of sparse tensor algebra workloads, we show that our proposed overbooking strategy introduces an average speedup of $52.7\times$ and $2.3\times$ and an average energy reduction of $22.5\times$ and $2.5\times$ over an existing accelerator without and with optimized tiling (\ie, uniform shape tiling and prescient uniform shape tiling), respectively. 
\end{itemize}

\section{Background}

\begin{figure*}[tb]
    \centering
    \includegraphics[width=\textwidth]{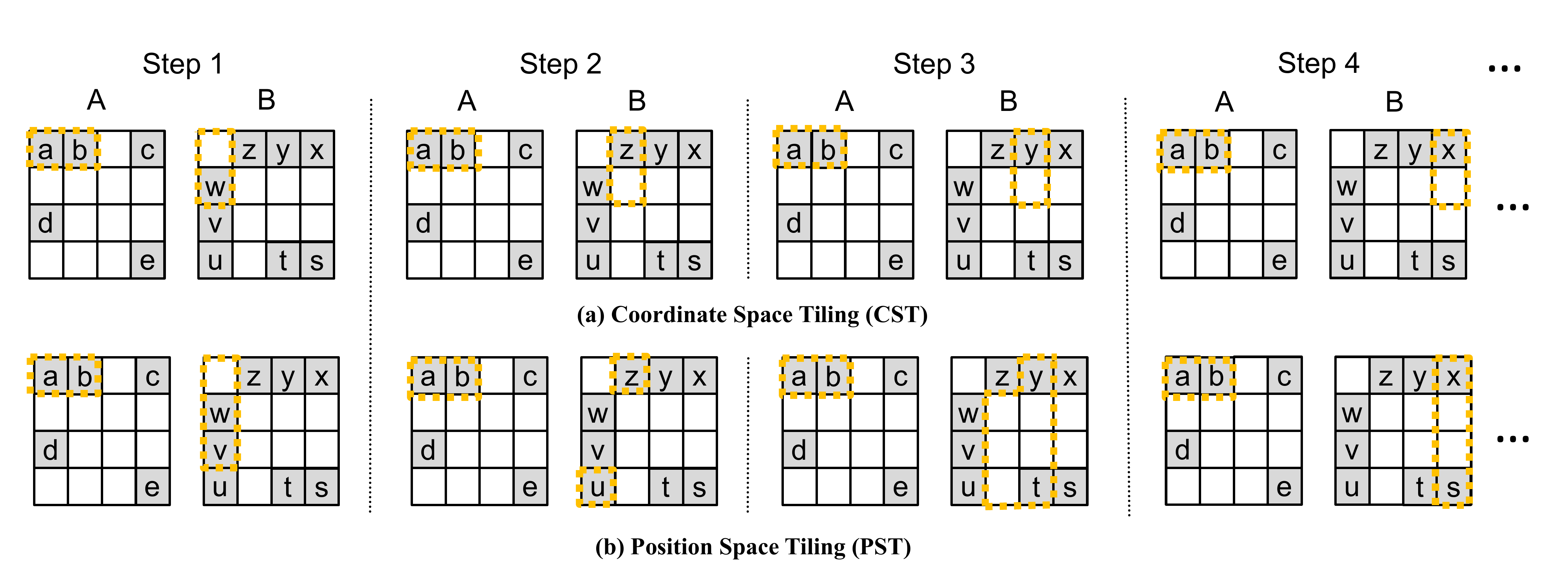}
    \vspace{-24pt}
    \caption{\textbf{Tiled sparse matrix multiplication between sparse 2-dimensional tensors (\ie, matrices) $A$ and $B$, when tiling in (a) coordinate space and (b) position space for a buffer with a capacity of two for each operand. Each step shows the tiles operated on.  Dotted yellow boxes indicate the tile in coordinate space. CST constructs $A$ and $B$ tiles with uniform shapes and thus does not require runtime operand matching. PST constructs $A$ and $B$ tiles of uniform occupancy, but can have potentially different shapes. Thus, PST  requires a costly runtime traversal of $B$ both to determine its tiling and to search for all possible matching operands given a tile from $A$.}}
   \label{fig:tiling:a}
   \label{fig:tiling:b}
\end{figure*}

This section discusses the basics of sparse tensor algebra, the limitations of prior tiling approaches, and various data orchestration approaches. 

\subsection{Sparse Tensor Algebra}
Tensors are multi-dimensional arrays of data, and when there exist zero values in the data, we call the tensor a \emph{sparse tensor}. Adopting the terminology from~\cite{sze_efficient_2020}, the logical locations of each element in a tensor, called \emph{points}, can be described by a tuple of \emph{coordinates}, one for each dimension. For example, for a two-dimensional tensor (\ie, a matrix), each data point can be defined by a (row, column) tuple. The tensors we focus on have integer coordinates; thus, the \emph{shape} of a tensor is described by a tuple of integer ranges and the \emph{size} of a tensor by the product of the ranges. 

Sparse tensor algebra involves applying various mathematical operations (\eg, multiplications and additions) on the data in multiple sparse tensors and can be described compactly with Einstein summation (Einsum) notation~\cite{einsum, sze_efficient_2020}. For example, matrix multiplication between a $M\times K$ tensor $A$ and a $K\times N$ tensor $B$ can be described as:
\begin{equation}\label{eq:einsum}
Z_{m,n} = A_{m,k} B_{k,n}
\end{equation}
This defines matrix multiplication for each point $(m,n)$ of the output as the sum of the products of elements of row $m$ in $A$ and column $n$ in $B$. Since sparse tensors introduce a significant number of \emph{ineffectual computations} (\eg,  $x\times 0 = 0$), many \emph{sparse tensor accelerators}~\cite{hegde_extensor_2019, GAMMA-accelerator, outerspace, sigma, matraptor, odemuyiwa_drt_2023} have been proposed to eliminate hardware operations associated with such ineffectual computations to improve hardware efficiency.

\subsection{Tiling Sparse Tensors}

To increase the arithmetic intensity for sparse tensor algebra processing, sparse tensor accelerators are often designed to have a multi-level memory hierarchy, and employ various tiling strategies to partition the tensors into \emph{tiles}, which are transferred to the next level (\ie, the level with buffers that have smaller capacities) in the memory hierarchy for data reuse. For sparse tensors, tiling becomes challenging as the exact occupancy of each tile (\ie, number of nonzeros) often cannot be determined without preprocessing or significant runtime processing. Moreover, sparsity can vary across the tensor and thus across equally-sized tiles of the tensor. 
In this section, we first formalize the tiling concepts introduced in Section~\ref{sec:intro} and discuss their limitations. 

In addition to their original form with both zeros and nonzeros, referred to as the uncompressed format,  sparse tensors can also be represented with compressed formats with only the nonzeros.  Thus, given a buffer with a certain capacity, compressed formats allow a larger tile to be stored than if the tile is uncompressed.  Adopting the terminology from~\cite{sze_efficient_2020}, we can classify tiling strategies as either:

 \begin{itemize}
     \item \textbf{Coordinate Space Tiling (CST)}: construct tiles with uniform shapes in the uncompressed space.
     \item \textbf{Position Space Tiling (PST)}: construct tiles with uniform occupancies based on the range of nonzero elements' positions in the buffers independent of each tile's shape.
 \end{itemize}
 However, we make the observation that neither of the above tiling approaches are simultaneously adaptable and efficient.

\subsubsection{Exploiting Sparsity with Coordinate-Space Tiling Requires Expensive Preprocessing}

CST tiling strategies~\cite{hegde_extensor_2019, matraptor} partition workload tensors into tiles of uniform shape. Specifically, a conservative CST approach partitions the workload assuming dense tensors. In this scenario, CST constructs tiles of a uniform fixed shape with a size that will always fit in the available buffers, independent of workload sparsity characteristics (\eg,  as indicated by the orange dotted boxes in Fig.~\ref{fig:tiling:a}a, given a buffer capacity of two, the tile size will always be two). Such a fixed tile shape allows the hardware to easily locate the corresponding tiles in other operands to perform the computations, (\ie, easy runtime operand matching). However, tiling sparse tensors with the assumption of dense tiles often leads to low buffer utilization, thus limiting data reuse (\eg, the buffers for \textit{B} are never fully utilized for the steps shown in Fig.~\ref{fig:tiling:a}a).  

The conservative CST approach can be enhanced to take tensor sparsity into consideration by partitioning the tensor into the largest possible tiles while guaranteeing each tile fits within the buffer. Achieving this requires traversal over the entire tensor for every possible tile shape to determine whether the tile with the largest occupancy still fits within the buffer. This compute-intensive step can be done during preprocessing for static data (\ie, the tensor is known \emph{a priori})~\cite{hegde_extensor_2019, odemuyiwa_drt_2023, jstream}; however, for input tensors generated by previous computation, this is done during runtime. Given a workload tensor, finding a good tile shape that will lead to higher buffer utilization typically involves expensive optimization approaches such as deep neural networks~\cite{waco} or inspector-executor schemes~\cite{mkl}. However, although taking maximum tile occupancy into account can potentially lead to higher buffer utilization by allowing tile shape to scale with sparsity, the significantly varying tile occupancy within a tensor can still result in a conservative tile shape, and thus low buffer utilization for most of the tiles.

\noindent\textbf{Takeaway: CST allows for easy runtime operand matching, but can have low buffer utilization even with heavy pre-processing.}

\subsubsection{Position-Space Tiling Requires Expensive Runtime Operand Matching}

PST allows high buffer utilization by partitioning the workloads into tiles with an occupancy that is equal to the available buffer capacity. Fig.~\ref{fig:tiling:b}b shows an example of performing PST on $A$ and $B$. For each processing step, given a buffer with a capacity of two, PST constructs tiles with two nonzero values whenever possible (\eg, $a$ and $b$ in $A$). 

However, even though PST is able to have high buffer utilization, PST incurs a high tiling tax since it needs to perform expensive runtime operand matching between sparse operand tiles. Specifically, since the distribution of nonzero value locations varies within and across tensors, with one operand tile constructed, PST needs to traverse over tiles of varying shapes in other tensors to search for all possible matching operands. For example, as shown in Fig.~\ref{fig:tiling:b}b, in order to locate the corresponding operands in $B$ for the nonzeros $a$ and $b$ in $A$, PST traverses tiles of shapes 3-by-1 at \emph{Step 1} and even across columns at \emph{Step 2}, resulting in a much more costly traversal compared to the CST example. Please note that since the tiles in $A$ can end up with arbitrary shapes, $B$ cannot be tiled apriori and PST always incurs the cost of full $B$ traversal for each tile of $A$. We make the observation that existing work that attempts to tile multiple sparse tensors in position space uses expensive control flow schemes and complicated tile management to build tiles~\cite{odemuyiwa_drt_2023}.

\noindent\textbf{Takeaway: PST allows high buffer utilization at the cost of complex/expensive hardware support for runtime operand matching.}

\subsection{Data Orchestration for Tiling}

Caches are commonly used as a buffering idiom for data orchestration in general-purpose computing (\eg, CPUs and GPUs). Assuming an optimal cache replacement policy, caches are able to manage tiles with occupancy greater than the cache size. However, caches incur high overhead for tag matching and associativity and are not typically suitable for accelerators~\cite{pellauer_buffet_2019}.

Another approach that is better suited for domain-specific accelerators is to perform \emph{explicit decoupled data orchestration (EDDO)}, where data movement is decided \emph{explicitly} by the program configuration and data requests are \emph{decoupled} from execution on the data~\cite{pellauer_buffet_2019, desc, stream-dataflow}. However, such techniques often have assumptions that are not friendly to sparse tensor algebra workloads. For example, buffets~\cite{pellauer_buffet_2019} are an EDDO storage idiom that features efficient decoupling of fine-grained synchronization and hierarchical composability, which are important attributes to have for domain-specific accelerator designs. However, the buffets idiom has a fixed assumption of the data reuse distance that can lead to poor reuse for sparse tensor algebra workloads. As a result, it is insufficient for efficiently utilizing available on-chip memory capacity.  

\noindent\textbf{Takeaway: Existing data orchestration approaches either introduce high control overhead or low buffer utilization for sparse tensor algebra workloads.}
\section{Hardware for Overbooking} 
\label{sec:overbooking_support}

In this section, we describe the concept of overbooking buffers and implement support for overbooking as an EDDO scheme for buffers. We first explain why existing EDDO approaches are insufficient for managing overbooked buffers and instead propose a hardware storage mechanism, called \emph{\obsb}, which efficiently support overbooking with low overhead. We then describe how tiles can be constructed to control for the degree of overbooking in Section~\ref{sec:overbooking_strategy}.

\subsection{General Concept}
Overbooking describes a strategy where tiles are allocated to a given buffer such that tiles have greater occupancy than the available buffer capacity (\ie, tiles may not fit in the buffer). This is achieved by speculating on the occupancy of tiles to determine whether a given tile will fit within the buffer. However, unlike traditional speculation schemes where ideal speculation is always accurate, overbooking-based speculation relies on some tiles not fitting to allow for larger tiles to be constructed. Essentially, overbooking is \emph{intentionally overconfident} when it speculates and ideal overbooking does not have all tiles fit within the buffer. We define $y\%$ overbooking to be a tiling strategy that leads to $y\%$ of tiles having occupancy greater than the buffer capacity. We will discuss the specifics of our tiling strategy in Section~\ref{sec:overbooking_strategy}.

As shown in Fig.~\ref{fig:longtail}, most tiles within a tensor have low occupancy and tile occupancies have high variability. Because of this distribution of tile occupancies in sparse tensors, being less than 100\% confident that a tile will fit in a given buffer allows for larger tiles to be allocated to that buffer, increasing buffer utilization (\ie, decreasing blank space) and thus data reuse. Compared to other existing tiling strategies, which must guarantee that the worst-case tile occupancy fits within a given buffer, overbooking enables larger tiles by constructing tiles that occasionally exceed the worst-case tile occupancy. 

Overbooking introduces challenges for data orchestration. Notably, because a tile is not guaranteed to fit within the target buffer, there will always be a cost in terms of reduced data reuse for tiles that overbook the buffer and the magnitude of this cost will depend on the data orchestration approach. Although EDDO approaches are commonly used in domain-specific accelerators, existing EDDO solutions are insufficient for supporting overbooking memory access patterns. We will first introduce the basic concepts behind EDDO approaches and then we will demonstrate the challenges of enabling overbooking with EDDO.

\subsection{Explicit Decoupled Data Orchestration}
\label{sec:buffets}
Explicit decoupled data orchestration (EDDO) defines a class of data orchestration approaches where data placement/removal in a buffer is workload-controlled (explicit) and each buffer can run at its own rate using data pushed to it (decoupled). EDDO approaches are commonly used in domain-specific accelerators because of their low overhead, ability to leverage static workload knowledge, and hierarchical composability. Two common methods of implementing buffers in EDDO approaches are FIFOs and buffets~\cite{pellauer_buffet_2019}.

FIFOs are a traditional buffer organization that introduce low overhead while enabling simple synchronization and hierarchical composability. FIFOs achieve this by restricting the access order and replacement policy to be first-in first-out. These restrictions are unacceptable for tensor algebra accelerators as the tensor algebra dataflow requires multiple accesses within a tile of data. 

To remove the restrictions emplaced by FIFOs, the buffets~\cite{pellauer_buffet_2019} storage idiom manages data to support \emph{random accesses} into the buffer and workload-controlled removal of data from the buffer. This is achieved by supporting four storage operations: Fill, Read, Update, and Shrink. These operations are described below.

\textbf{Fill(Data):} Fill describes how a new element of \emph{Data} is written into the buffer. This is done by managing the buffer as a queue: with a known head pointer and a known buffer occupancy, new data is placed at the tail of the queue. 

\textbf{Read(Index):} Read describes how random accesses into data within the buffer are performed. Because buffets manage the buffer as a queue, reads are performed relative to the head of the queue and the \emph{Index} is used to refer to the offset from the head of the queue. Thus, index 0 represents the data at the head of the queue and the largest possible index is equal to the buffer capacity. When the index read exceeds the current buffer occupancy (\ie, the tail of the queue) the read stalls until the data arrives. 

\textbf{Update(Index, Data):} Update describes how elements within the buffer can be modified. While Fill and Update both write data into the buffer, Update is the only way to change the value of data inside the buffer and the only way to write to an arbitrary index within the buffer. Similar to reads, writes are performed relative to the head of the queue based on the \emph{Index}: the element at a given \emph{Index} is updated with \emph{Data}. By supporting read/write operations with indexing, buffets support random accesses into the buffer. 

\textbf{Shrink(Num):} Shrink describes how data is removed from the buffer. Within the queue abstraction, shrinks free data from the head of the queue by incrementing the head pointer by \emph{Num} to indicate the number of data elements to remove from the buffer and shrinking the occupancy by \emph{Num}. Synchronization between shrinks and fills is achieved using a credit system: data is only pushed to the buffet for fills when credits are available (\ie, credits indicate the number of unoccupied slots in the queue). Following a shrink, \emph{Num} credits are released to indicate that another fill can be performed with the newly freed occupancy of the buffer. 

With these four operations, buffets are able to support random access to any data held within the buffet. Buffets utilize a queue abstraction to store data within a buffer to enable simple management and synchronization; thus, they are limited to data access patterns that behave as a sliding window over the data (\ie, fill from the head, shrink from the tail). A sliding window-based data removal pattern is insufficient for overbooking due to the lack of fine-grained control over what data can be removed from the buffer. 

When a tile overbooks a given buffer, some data within the tile cannot fit within the available buffer capacity. We refer to this data as \emph{bumped data}.\footnote{A bumping incident on an airline occurs when a flight is overbooked and too many passengers attempt to board. Typically, a bumped individual is provided substitute transportation and monetary compensation. \obsb~provide no compensation because nonzeros do not have legally protected rights, but Tailors do provide alternative "transportation", so one can be fearless when overbooking without it becoming treacherous.} Using existing tiling strategies with existing data orchestration methods such as buffets, the entire tile fits within the buffer and it is possible to exploit data reuse within the buffer. However, when the buffer is overbooked and data within the tile is bumped, data reuse is lost. 

The problem with supporting overbooking with buffets is that \emph{buffets can only free the oldest data held within the buffer} (\ie, shrink from the head). We show how buffets manage an overbooked tile in Fig.~\ref{fig:osb-buffet-comparison}. The sliding window that the buffet operates on has length 3, which is shorter than the reuse distance of 4 of the data, causing the buffet to remove data ($v$, $w$, $x$, $y$) that ends up being reused in the future. When the sliding window that the buffet wants to operate on is larger than the buffer, the buffet has no choice except to drop everything and re-fill the full tile each time it traverses the tile. In contrast, Tailors only need to re-fill overbooked elements within the tile.

\subsection{\obsb}

\begin{figure*}[t]
    \centering
    \includegraphics[width=\textwidth]{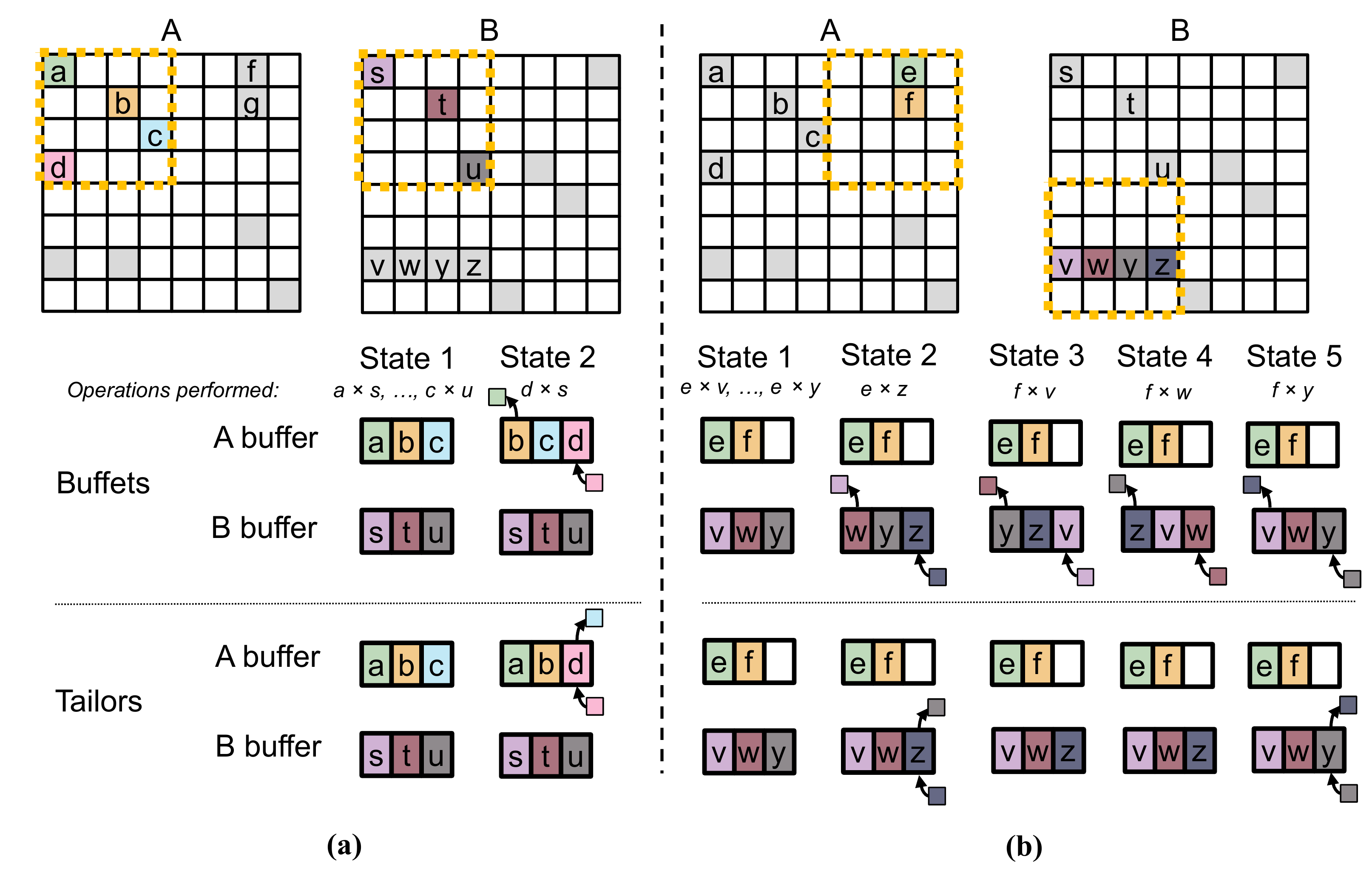}
    \vspace{-25pt}
    \caption{\textbf{Comparison of data management between \obsb~and buffets when (a) a tile from the stationary operand $A$ overbooks the buffer and (b) a tile from the non-stationary operand $B$ overbooks the buffer.  Nonzeros in each sparse tensor are shown with colour and the tiles needed for the computation are outlined by dotted yellow boxes. Each state describes the data residing in the buffer after the data in the buffer changes. Data is removed from the buffer when the buffer is full and an element not residing in the buffer is required for an operation.} Arrows are used to indicate data movement. An arrow into the buffer indicates data being written into the buffer, while an arrow out of the buffer indicates data being removed from the buffer. While the buffet continuously cycles data in the buffer, the Tailor is able to reuse a portion of the data.}
       \label{fig:osb-buffet-comparison}
        \phantomsubcaption
    \label{fig:osb-buffet-comparison:a}
    \phantomsubcaption
    \label{fig:osb-buffet-comparison:b}
\end{figure*}

To address the limitations of buffets, we develop Tail Overbooked Buffers, or \emph{\obsb}, as an extension of buffets to enable data reuse even when a tile does not fit within the buffer. Specifically, we handle bumped data by (repeatedly) streaming the bumped portion of the tile through the buffer before we must begin again. To support this, we overwrite a fixed space at the tail of the buffer when the buffer is full and use that space for streaming. This approach ensures that most data held within the buffer is not bumped to satisfy the requests for new data in a given tile. As a result, \obsb~are still able to exploit data reuse for a portion of an overbooked tile. We show how \obsb~manage data for an overbooked tile in Fig.~\ref{fig:osb-buffet-comparison}. \obsb~explicitly manage streaming to only remove data from a fixed space at the tail of the buffer, only overwriting data that is used for streaming. Thus, in State 3 and 4 of Fig.~\ref{fig:osb-buffet-comparison:b}, \obsb~are able to reuse data already in the buffer to complete the operation (\ie, all 'v' and 'w' have to do is stay in the buffer). 

\begin{figure}[t]%
    \centering
\includegraphics[width=\columnwidth]{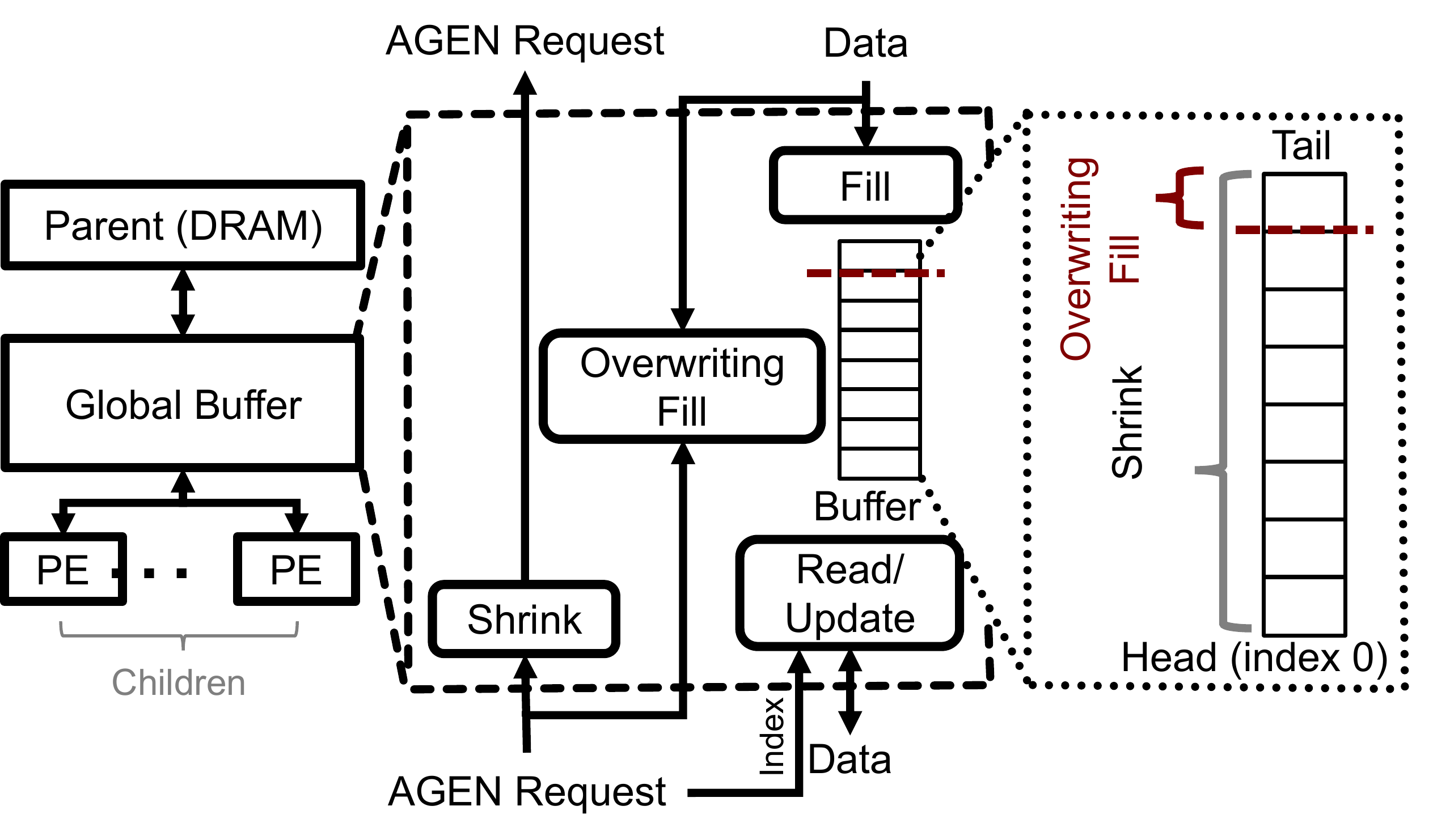}
\vspace{-20pt}
    \caption{\textbf{(Left) A typical accelerator memory hierarchy made up of global buffers, PE buffers, and compute in each PE. Each buffer is associated with an address generator (AGEN) which generates addresses for future fills. (Middle) \obsb-defined operations on the buffer. (Right) Where data can be freed from the buffer for a given operation. Overwriting fills only modify the tail of the buffer when the buffer is full, while shrinks can modify the entire buffer starting from the head, and fills can modify the buffer when it is not full. }}
   \label{fig:hierarchy}
\end{figure}

\obsb~provides a scan resistance similar to the Bimodal RRIP~\cite{rrip} cache replacement policy, however, rather than being cache based \obsb~can be integrated into memory hierarchies as an EDDO storage idiom. To illustrate the idea, without loss of generality, we use the accelerator architecture organization in Fig.~\ref{fig:hierarchy} as an example. The example architecture organization consists of multiple memory levels, with the DRAM as the highest and buffers in the PEs as the lowest. We refer to the parent of any memory level as the memory level above it and the child as the memory level below it (\eg, in Fig.~\ref{fig:hierarchy}, the parent of the global buffer is DRAM, while the children of the global buffer are the PEs). Each buffer is controlled by a sparse address generator (AGEN), which traverses the compressed representation of a tile to push data to its children. Shrinks are driven by the child buffer's address generator, fills are driven by data from the parent (DRAM), and reads/updates interface with the children (PEs). To support streaming through the buffer when overbooked, we free space at the tail of the buffer to use for data streaming and overwrite existing data with the data needed to provide fills for the child. Because we only modify a small portion at the tail of the buffer, the rest of the data (close to the head) can stay in the buffer for reuse.

\subsubsection{Realization of Tail Overbooking} We realize tail overbooking by implementing a streaming interface for the buffets EDDO scheme. To stream data through a buffer, the intuitive solution is to have a separate FIFO for bumped data to pass through. However, this solution does not bring us out of the woods since it requires additional on-chip memory that could instead be used to store larger tiles. Instead, \obsb~support FIFO-like operation by extending the buffet interface with an additional modified fill operation, the \emph{overwriting fill}, which is used to overwrite data at the tail of the buffer. This enables deque-like management of data in the buffer and allows for the tail of the buffer to be used for streaming data while the head is used for general buffer management.

Essentially, Tailors have two modes: (1) when a tile completely fits within the buffer and the buffer is not overbooked, it allows the buffer to be managed as a buffet; (2) when a tile does not fit within the buffer and overbooks the buffer, Tailors partition the buffer into a buffet-managed region as described in Section~\ref{sec:buffets} and a FIFO-managed overbooked region that is managed with overwriting fills instead of the general buffet fills. To keep the impact of an overwriting fill local, an overwriting fill is limited to affecting the tail of the buffer (\ie, the FIFO-managed region). The size of the FIFO-managed region used for streaming in the buffer is configurable. If this space is too small, data streaming may bottleneck execution due to the latency of sending data to children. However, if this space is too large, data reuse in the buffer is reduced as data that could have been reused is removed to fit streamed data. We statically set the size of the FIFO-managed region such that the round-trip latency between the buffer and its parent can be hidden by double-buffering and thus avoid bottlenecking child buffers (\ie, same partitioning for all workloads); however, another possible solution to this problem is to partition the regions at runtime and adapt to whether execution is memory-bound, so our static solution is not the endgame. 

At a high level, overwriting fills have the same interface as fills: \textbf{OWFill(Data)} writes \emph{Data} to the tail of the buffer. However, unlike conventional fills, overwriting fills atomically shrink from the tail of the buffer to accommodate \emph{Data} fill rather than decoupling the shrink from the fill. These overwriting fills operate in the FIFO-managed region of the buffer. We discuss how Tailors manage data in this section and provide an example of overbooking with Tailors in Section~\ref{sec:extailors}. 

When a Tailor sees an initial overwriting fill, it clears the space of the FIFO-managed region by atomically clearing part of the buffet-managed region by the size of the FIFO-managed region and filling the region with the bumped data from the tile. Subsequent overwriting fills modify the FIFO region of the buffer without touching data in the buffet-managed region. Thus, accesses to data held in the buffet-managed region can continue to be reused without any additional cost.

\subsubsection{Maintaining Support for Buffet Semantics}
Maintaining support for the original buffet semantics within Tailors enables efficient data orchestration. In this section, we discuss how Tailors maintain support for the various buffet operations.

\textbf{Maintaining support for Fill:} Streaming support within \obsb~is achieved using the overwriting fill operation, which cannot be followed by fill operations as both write to the tail of the buffer. Allowing both to happen at the same time would introduce race conditions which lead to loss of data since the data that was written over by an overwriting fill is removed from the buffer and there is no mechanism to easily recover it. 

\obsb~avoid such race conditions by mandating that streaming -- and thus the use of overwriting fills -- only occurs when the buffer is full, which naturally blocks fills based on original buffet semantics. Moreover, to support streaming, the space in the buffer that the overwriting fill overwrites is kept the same so long as no shrink is performed.

\textbf{Maintaining support for Read/Update:} Writing to the tail introduces a key difference between Tailors and buffets: while in buffets the \emph{Index} (\ie, the location in the current tile) and the \emph{Offset} (\ie, the location in the buffer) are identical because data is managed as if it were a contiguous sliding window, this is not true for Tailors since Tailors can divide the buffer into separate buffet-managed and FIFO-managed regions. 

To maintain the sliding window abstraction and thus compatibility with buffets, Tailors track the difference between the \emph{FIFO head} (\ie, the start of the FIFO-managed region) and the index of the least recent data in the buffer. We call this difference the \emph{FIFO offset}. Similarly, we use the terms \emph{buffer head} and \emph{buffet offset} to indicate the start of the buffer (\ie, always zero) and the location in the buffer, respectively. Given an initial overwriting fill, the FIFO offset is set to be equal to the size of the FIFO-managed region. Whenever an overwriting fill replaces earlier data, Tailors increments this value by one. The FIFO offset is reset to zero when a data read occurs to data in the buffet-managed region of the buffer. With this FIFO offset, it then becomes possible for reads and updates to index into the buffer without modification of read semantics even when some data has been bumped. This is done by subtracting the FIFO offset from the Index (\ie, \emph{Index} - \emph{FIFO Offset}) to get the position from the head of the queue to access. 

To divide a buffer into two regions, Tailors defines a head pointer to indicate the start of each region. Although we implement buffer management with a rolling buffer, we discuss offsets and heads as though they are fixed for simplicity. For the buffet-managed region, the head always points to the start of the buffer (\ie, an offset of 0). In contrast, the FIFO head points to the start of the FIFO-managed region and is equal to the size of the buffet-managed region.
To determine whether to index using the FIFO offset or not, Tailors compares the index to the two head pointers. For indices less than the difference between the FIFO head and the buffet head, accesses go to the buffet-managed region and can use the index directly as the offset into the buffer. For indices greater than the difference, accesses go to the FIFO-managed region, and the FIFO offset is needed to compute the offset into the buffer.

\textbf{Maintaining support for Shrink:}
When a shrink occurs and frees data from the head of the buffer, the buffer will no longer be full and, if overwriting fills continue, buffer utilization will be reduced. Thus, a shrink triggers overwriting fills to backfill the buffer with the tile that caused the buffer to be overbooked. To maintain coherent indexing, backfill only occurs after reaching data held in the buffet-managed region of the buffer. If the buffer still cannot hold the tile and is overbooked, the remaining bumped data continues to be handled by overwriting fills. If the buffer is no longer overbooked, the parent can push new data to the buffer as credits will be available. 

By only modifying the interaction between the parent and the buffer itself, \obsb~maintain the hierarchical composability of prior EDDO schemes and enables the hierarchical integration of \obsb~into memory systems.

\subsubsection{Example of Overbooking with Tailors}
\label{sec:extailors}

\begin{figure}[t]%
    \centering
\includegraphics[width=\columnwidth]{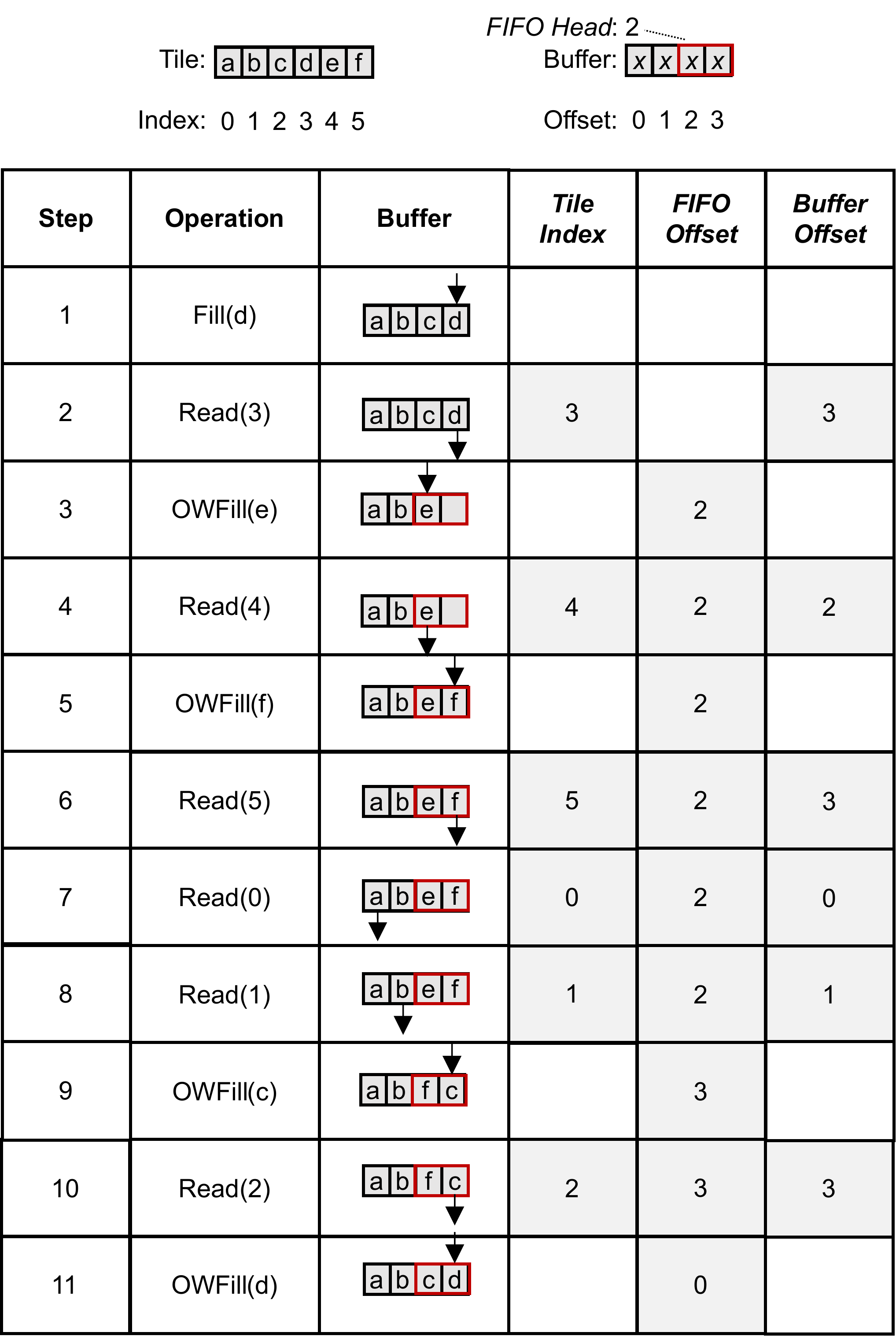}
\vspace{-16pt}
    \caption{\textbf{\obsb~management following an example sequence of consecutive operations caused by overbooking with a buffer that can hold four elements. The FIFO-managed region is configured to hold two elements. Red boxes indicate the FIFO-managed region of the buffer and arrows indicate data movement. Arrows into the buffer indicate data fills from the parent, while arrows out of the buffer indicate data sent to the child. The \emph{FIFO Offset} (\ie, the difference between the \emph{FIFO Head} and the index of the least recent data in the FIFO) and the \emph{Buffer Offset} (\ie, the location in the buffer) used to index into the buffer are shown. We implement the FIFO-managed region as a rolling buffer with a head pointer but show it with a fixed head position for simplicity.  }}
   \label{fig:overwriting-fill}
\end{figure}

Fig.~\ref{fig:overwriting-fill} illustrates a sequence of operations with \obsb~and illustrates how Tailors tracks data over the course of operation on an overbooked tile. Following the \textbf{Fill(d)} operation, the buffer becomes full while there is still data in the tile. Thus, the initial overwriting fill \textbf{OWFill(e)} splits the buffer into a buffet-managed region and a FIFO-managed region (outlined in red). Since the Tailor was configured with a FIFO-managed region of size two, the FIFO offset is set to two and the FIFO head is also set to two. 

With the subsequent \textbf{OWFill(f)} operation, the FIFO-managed region is full. The \textbf{Read(5)} operation accesses index 5 in the tile. Since this accesses the FIFO-managed region, the buffer offset read from the buffer is 3 (\emph{Buffer Offset} = \emph{Index} - \emph{FIFO Offset}).

Since the following data reads (\textbf{Read(0)} and \textbf{Read(1)}) are from indices less than the FIFO head, they proceed without modification. However, subsequent overwriting fills must select some data to replace. Since overwriting fills operate solely on the FIFO-managed region, the following \textbf{OWFill(c)} operation drops the oldest data (\textbf{e}) and increments the offset by one. Due to the rollover of data (\textbf{c}), the \textbf{Read(2)} operation rolls over indexing and thus accesses an offset of 3.

The operation that follows (\textbf{OWFill(d)}) replaces the data at the end of the tile (\textbf{f}) and thus resets the FIFO offset to zero.

\section{Overbooking Tiling Strategy}
\label{sec:overbooking_strategy}

In this section, we describe an adaptable and efficient tiling strategy, \emph{Swiftiles}, to construct coordinate-space tiles (CST) that may overbook a given buffer. 
\begin{figure*}
    \centering
    \includegraphics[width=\textwidth]{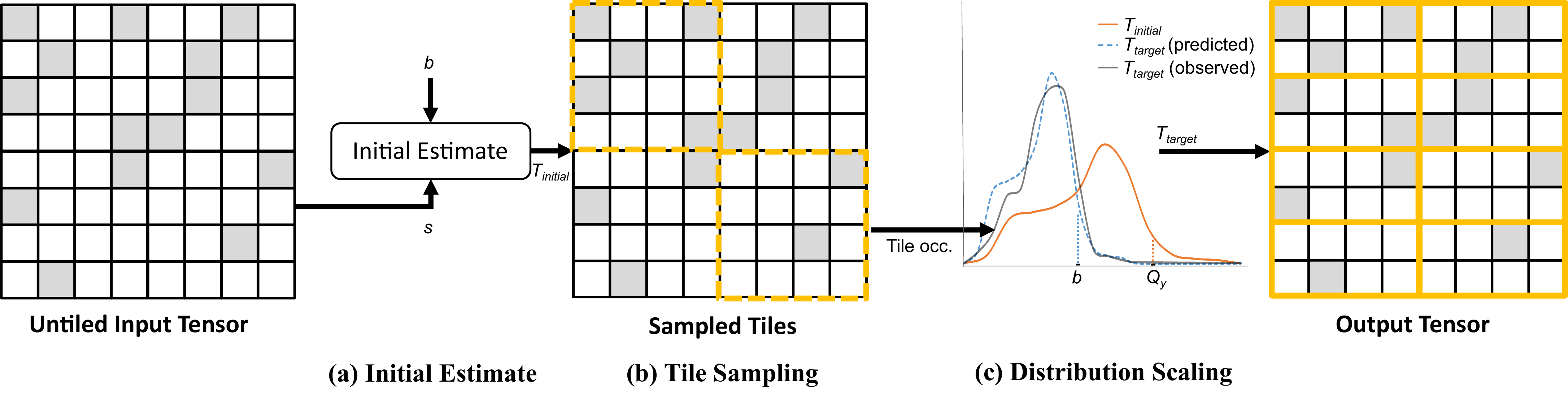} 
    \vspace{-24pt}
    \caption{\textbf{Overview of Swiftiles operating on a sparse tensor. Darker squares show nonzeros while white squares show zeros. Dotted yellow boxes are used to show sampled tiles and solid yellow boxes are used to show the final tiling. (a) Initial estimate $T_{initial}$ is constructed using global average sparsity $s$ of the tensor and buffer capacity $b$, (b) Tiles are sampled from the tensor using the initial estimate to generate a list of tile occupancy samples, (c) The tile occupancy distribution when the tensor is tiled using different tile sizes. After tiling using the initial estimate $T_{initial}$ to generate the sampled distribution (shown in orange), Swiftiles finds the $y\%$ quantile (shown with ellipses), and the distribution is scaled so that the $y\%$ quantile fits exactly inside the buffer. This gives the resulting predicted distribution with Swiftiles (shown in blue) and the final prediction $T_{target} \text{(predicted)}$. We show the observed tile occupancy distribution (\ie, the distribution obtained by traversing the entire tensor at the given tile size) for when the tensor is tiled with $T_{target}$ in black as $T_{target}\text{(observed)}$.
    }}
    \label{fig:swiftiles}
    \phantomsubcaption
    \label{fig:swiftiles:a}
    \phantomsubcaption
    \label{fig:swiftiles:b}
    \phantomsubcaption
    \label{fig:swiftiles:c}
\end{figure*}

\subsection{Preprocessing for Tile Construction}
The efficacy of overbooking depends on the frequency of the buffer being overbooked by a tile. In overbooking, this is described by a confidence threshold $y$ on which $y\%$ of tiles will overbook the target buffer (\ie, $y\%$ equals the number of tiles that are overbooked out of the total number of tiles). However, determining the exact tile size necessary to minimize memory traffic for any given confidence has a prohibitive preprocessing cost of checking the tile occupancy of each tile for all possible tile sizes. 

For example, prescient CST can be framed as 0\% overbooking where no tiles overbook the buffer. Determining whether a given tile size never causes a buffer to overbook on a given tensor requires fully traversing the tensor to compute the tile occupancies of each and every tile for a given tile size. Thus, to determine the \emph{maximum tile size} that never overbooks, this traversal must be done across a huge number of candidate tile sizes, resulting in a preprocessing cost that scales with the size of the tensor and the number of candidate tile sizes. This cost can easily dominate the cost required to perform the actual sparse tensor operation. 

As a result, it is necessary to have a tile construction technique for arbitrary $y$ which minimizes construction cost and, ideally, decouples the cost of preprocessing from the size of each tensor.

\subsection{Swiftiles}
We propose an adaptable and efficient tile size search strategy, \emph{Swiftiles}, to swiftly size tiles for a given confidence. Swiftiles targets a confidence $y>0$ for a given tensor and tries to select a tile size where $y\%$ of tiles lead to overbooking in the buffer.

To minimize preprocessing cost, Swiftiles performs tile size estimation using a one-shot sampling scheme separated into three steps: (1) Swiftiles makes an \emph{initial estimate} of the tile size $T_{initial}$ without traversing the tensor. (2) Swiftiles performs \emph{tile sampling} using this tile size to create a sampling distribution of tile occupancies using samples of tiles from the tensor. This tile occupancy distribution aims to capture variability in sparsity between tiles of the tensor. (3) By assuming that small changes in tile size do not significantly change the shape of the distribution, Swiftiles \emph{scales the distribution} so that the $y\%$ quantile fits within the buffer and produces the final prediction $T_{target}$. We evaluate the change in distribution caused by a change in tile size in Fig.~\ref{fig:scaling} and show an example of this distribution shift in Fig.~\ref{fig:amazon0312}.

Swiftiles optimizes for \emph{tile size} rather than \emph{tile shape} because tile shape is often dependent on the dataflow and estimation based only on tile size is more tractable. Fig.~\ref{fig:swiftiles} shows a general overview of how Swiftiles estimates the tile size for a given confidence threshold. We discuss the three steps of Swiftiles in detail in the following sections.

\subsubsection{Initial Estimate $T_{initial}$}
In Swiftiles, an initial estimate of the tile size is used to partition the target tensor for sampling (Fig.~\ref{fig:swiftiles:a}). Since the degree of variability in the tile occupancy distribution depends on tile size, the tile size used when constructing tiles for sampling is important for ensuring the reliability of the sampling distribution. Generally, smaller tile sizes have greater variability due to capturing more fine-grained detail in the sparsity pattern, while larger tile sizes have less variability due to averaging over a larger number of elements. 

The initial estimate has two key design considerations: (1) To minimize preprocessing cost, the initial estimate should be computable in constant time. (2) Since Swiftiles makes the reasonable assumption that small changes in tile size do not significantly affect the shape of the tile occupancy distribution, $T_{initial}$ should also scale proportionally to $T_{target}$ and be roughly close to $T_{target}$. To meet both considerations, Swiftiles uses the tensor average sparsity $s$ and the buffer capacity $b$ to construct $T_{initial}$:
\begin{equation}
T_{initial} = \frac{b}{1-s}.
\end{equation}
The tensor average sparsity can be computed using only the shape of the tensor and the total number of nonzeros in the tensor, values that are typically available without having to traverse the tensor. In the overbooking framework, this estimate would describe the tile size needed for $50\%$ overbooking (\ie, confidence threshold of 50\%) when nonzeros are uniformly distributed across the tensor. Notably, $T_{initial}$ scales with the tensor size and sparsity, although not necessarily with the variability of sparsity between tiles nor the value of $y$. These variations are captured and corrected in the later steps of Swiftiles.

\subsubsection{Tile Sampling}
Using the initial estimate, Swiftiles tiles the tensor and samples the tile occupancy of different tiles in the tensor (Fig.~\ref{fig:swiftiles:b}). If all tiles are sampled, Swiftiles produces the exact tile occupancy distribution at $T_{initial}$. However, because iterating over the entire tensor to sample all the tiles is expensive, Swiftiles adopts a random sampling strategy that uses a fixed number of samples depending on the confidence threshold $y$. 

Specifically, Swiftiles selects $k$ as the number of samples that fall in the top $y\%$ quantile of sampled tile occupancies. This ensures that, regardless of what $y$ is selected, Swiftiles is able to identify enough samples to make a good approximation of the true tile occupancy distribution. For example, for $y=10\%$, Swiftiles collects $\frac{k}{0.1}=10\times k$ samples to construct the sampling distribution. We statically set $k$ and leave the per-workload selection of $k$ based on the tensor to future work. We show the results of a sweep of sampling choices in Section~\ref{sec:swiftiles-parameters}.

\subsubsection{Distribution Scaling}
Following tile sampling, Swiftiles has a sampling distribution of tile occupancies for when the tensor is tiled using the tile size $T_{initial}$, which is scaled to make the final prediction (Fig.~\ref{fig:swiftiles:c}). Swiftiles then finds the $y\%$ quantile point $Q_y$, which is the point that $y\%$ of sampled tiles have occupancy greater than. However, $Q_y$ does not consider the buffer capacity and how many tiles would overbook the actual buffer. To adjust to the actual buffer capacity, Swiftiles scales $T_{initial}$ using the point $Q_y$ and the capacity of the target buffer $b$ to get $T_{target}$: 
\begin{equation}
T_{target} = T_{initial} \times \frac{b}{Q_Y}.
\end{equation}
This linear scaling to produce the final prediction $T_{target}$ from $T_{initial}$ assumes that the tile occupancy distribution between $T_{initial}$ and $T_{target}$ are strongly correlated. As shown in Fig.~\ref{fig:swiftiles:c}, the scaled distribution ($T_{target}$ (predicted)) may still differ from the observed distribution ($T_{target}\text{(observed)}$): Swiftiles aims to minimize the difference between these two distributions at the $y\%$ quantile point. We show that this assumption is accurate in Fig~\ref{fig:scaling}. With this correlation, Swiftiles is able to make accurate predictions of the tile size needed for $y\%$ of tiles to overbook the buffer without measuring the tile occupancy distribution for different tile sizes, even if the distribution may not be identical.

\section{Methodology}

We integrate overbooking into the state-of-the-art CST-based ExTensor~\cite{hegde_extensor_2019} and evaluate over a set of sparse tensor algebra workloads. 

\subsection{Evaluation Platform}
We use the Sparseloop-Accelergy infrastructure~\cite{timeloop, sparseloop, accelergy} to model the various accelerator designs.  Sparseloop-Accelergy captures an accelerator’s cycle counts and component runtime activities. We implement a new sparsity model in Sparseloop to capture sparsity characteristics based on the per-tile data occupancy extracted from sparse tensors. We characterize energy consumption of various components using an Accelergy energy-estimation plug-in: 1) for datapath components, we used synthesized RTL with a 65nm PDK; 2) for small SRAMs, we used a 65nm SRAM compiler; 3) for large SRAMs, we used CACTI~\cite{cacti}.

\renewcommand{\arraystretch}{1.4}
\begin{table}[htbp]
    \small
    \centering

    \begin{tabular}{l|l|l}
        \textbf{Tensor} & \textbf{Dimensions} & \textbf{Sparsity} \\
        \hline
        rma10 & $47k\times 47k$ & $99.89\%$ \\
        cant & $63k\times 63k$ & $99.90\%$ \\        
        consph & $83k\times 83k$ & $99.913\%$  \\        
        shipsec1 & $141k\times 141k$ & $99.960\% $  \\
        pwtk & $218k\times 218k$ & $99.971\%$\\
        cop20k\_A & $121k\times 121k$ & $99.982\%$\\
        mac\_econ\_fwd500 & $207k\times 207k$ & $99.997\%$ \\ 
        mc2\_depi & $525k\times 525k$ & $99.9992\%$\\        
        \hline
        pdb1HYS & $36k\times 36k$ & $99.67\%$\\  
        sx-mathoverflow & $24k \times 24k$ & $99.96\%$\\
        email-Enron & $37k\times 37k$ & $99.973\%$\\        
        cage12 & $130k\times 130k$ & $99.988\%$ \\ 
        soc-Epinions1 & $76k\times 76k$ & $99.991\%$ \\
        soc-sign-epinions & $131k\times 131k$ & $99.995\%$\\
        p2p-Gnutella31 & $63k\times 63k$ & $99.996\%$\\  
        sx-askubuntu & $159k\times 159k$ &$99.997\%$\\
        amazon0312 & $400k\times 400k$ & $99.998\%$\\        
        patents\_main & $241k\times 241k$ & $99.999\%$\\
        email-EuAll & $265k\times 265k$ & $99.9994\%$ \\
        web-Google & $916k\times 916k$ & $99.99958\%$\\      
        webbase-1M & $1.0M\times 1.0M$ & $99.99968\%$\\
        roadNet-CA & $2.0M\times 2.0M $ & $ 99.99986\%$ \\        
    \end{tabular}
    \vspace{4pt}
    \caption{\textbf{Characteristics of the tensors used in the evaluation. Tensors listed in the top half are  constructed from systems of linear equations, while those listed in the bottom half are from other applications with sparse operands. Tensors are sorted by sparsity in decreasing order.}}
    \label{tab:benchmarks}
\end{table}
\renewcommand{\arraystretch}{1.0}

\begin{figure*}[t]%
    \centering
\includegraphics[width=\textwidth]{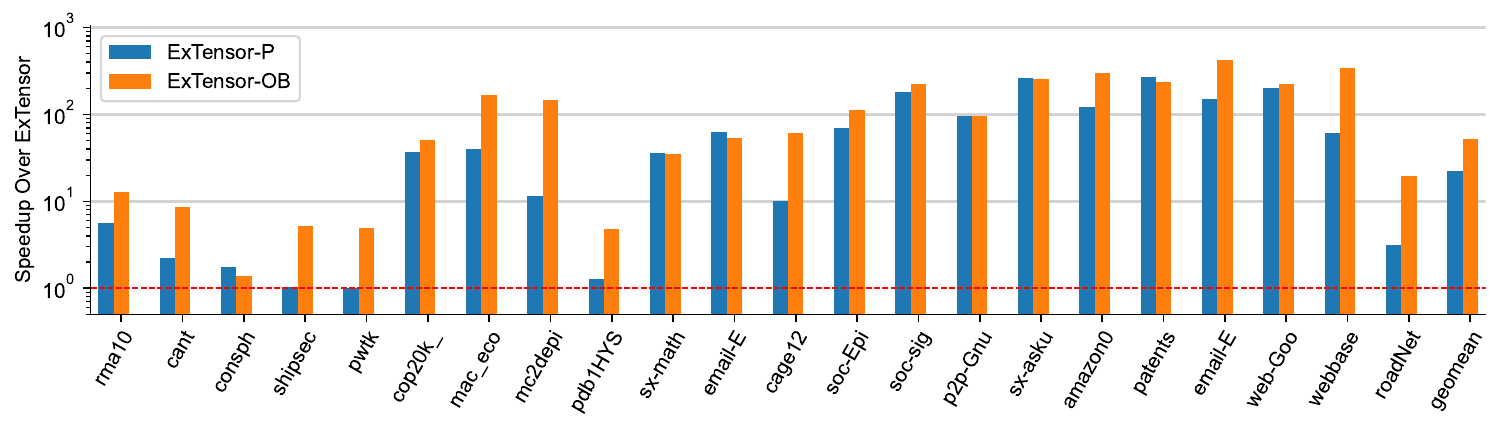}
    \vspace{-24pt}
    \caption{\textbf{ExTensor-P and ExTensor-OB speedup relative to ExTensor-N. ExTensor-N's performance is shown with a red line.  }}
    \vspace{-4pt}
   \label{fig:speedup}
\end{figure*}

\begin{figure*}[t]%
    \centering
\includegraphics[width=\textwidth]{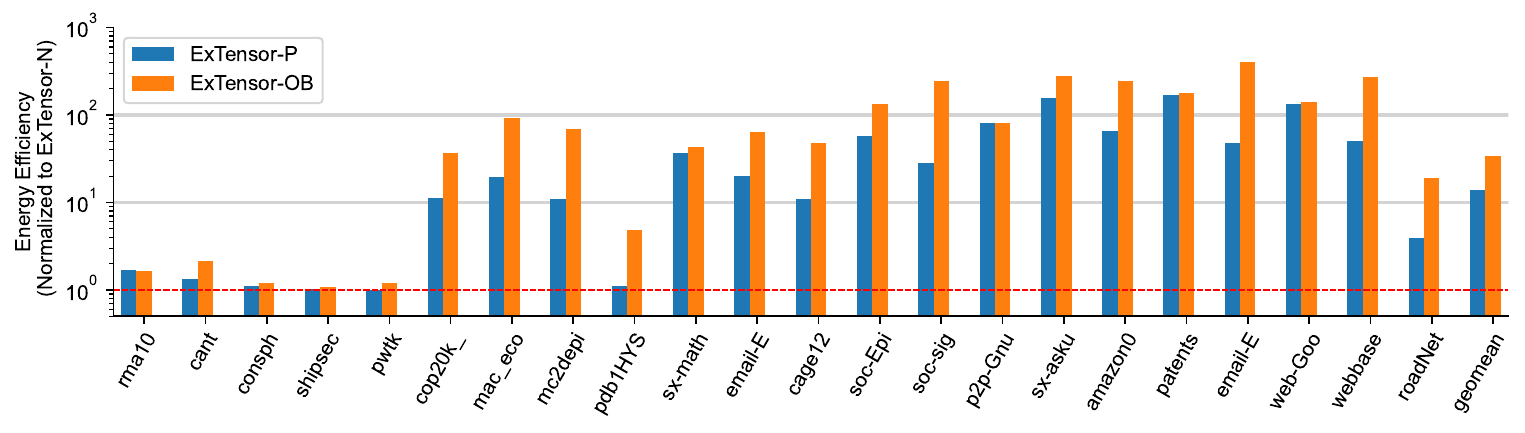}
    \vspace{-24pt}
    \caption{\textbf{ExTensor-P and ExTensor-OB energy relative to ExTensor-N. ExTensor-N's performance is shown with a red line. }}
    \vspace{-4pt}
   \label{fig:energy}
\end{figure*}

\subsection{Accelerator Designs}
\label{sec:acc-design}
We demonstrate how overbooking can improve upon ExTensor, a state-of-the-art sparse tensor algebra accelerator. ExTensor proposes to tile both operands in coordinate space and perform intersection between streams of nonzero coordinates. Coordinate streams are constructed by accessing coordinate metadata (which indicates where nonzeros occur in the tensor) for both operand tiles in a scan access pattern over the shared dimension. Operands are stored in compressed sparse fiber~\cite{sze_efficient_2020} format in separate buffers.

ExTensor uses CST along all three memory levels including DRAM, global buffer, and PE buffers. We evaluate three variants of ExTensor with different tiling strategies: the original ExTensor design without preprocessing (ExTensor-N), ExTensor with prescient preprocessing (ExTensor-P), and ExTensor with overbooking using \obsb~and Swiftiles (ExTensor-OB). The ExTensor-N uses fixed coordinate-space tile size and shape across all workloads to always fit within a given buffer, avoiding any preprocessing cost by constructing tiles according to the size of the tile rather than the occupancy. We extend ExTensor-N by creating a baseline ExTensor-P, which constructs tiles based on knowing the worst-case observed tile occupancy prior to tiling. Thus, ExTensor-P shows the performance of the best-possible CST without exceeding the size of any given buffer. In practice, ExTensor-P would incur a significant preprocessing overhead due to needing to check the occupancy of each tile at all tile sizes. We compare these two baselines to ExTensor-OB, which uses \obsb~to support overbooking and uses Swiftiles targeting 10\% overbooking to determine tile size. 

For 2D tensors, ExTensor-N uses fixed $128\times 128$ size tiles for PE buffers and sizes global buffer tiles to fit the worst-case occupancy of PE buffers (\ie, that each tile is dense). We construct tiles for ExTensor-P and ExTensor-OB by first expanding along the shared $K$ dimension between two operands until reaching the end of the dimension, then along the $N$ dimension for operand $B$, then along the $M$ dimension for operand $A$. This tile construction strategy maximizes output reuse given the original ExTensor dataflow. Similar to ExTensor-N, ExTensor-P and ExTensor-OB first partition a tensor into tiles for the global buffer, then partition the global buffer tile into subtiles for each of the 128 PE buffers. 

We normalize the configuration of all evaluated accelerators to that described in the original ExTensor paper at 1GHz. ExTensor uses a 30MB global buffer with 128 PEs and 4 DRAM channels with a total bandwidth of 68.25 GB/s. 

\subsection{Workloads}
We evaluate performance using real-world tensors from the SuiteSparse Matrix Collection~\cite{kolodziej_suitesparse_2019} spanning a range of sparsities, sparsity patterns, application domains, and tensor dimensions. We select tensors that span a wide range of sparsities and observe that tensors with high sparsity tend to have greater variation in tile occupancy. Similar to prior work~\cite{GAMMA-accelerator, hegde_extensor_2019, matraptor}, we evaluate SpMSpM computing $A\times A^T$. The tensors used in our evaluation are summarized in Table~\ref{tab:benchmarks}.

We note that a large majority of the tensors in SuiteSparse are built from large systems of linear equations. Systems of linear equations are typically represented as sparse 2D tensors with many nonzeros near the diagonal and few nonzeros away from the diagonal because of the nature of linear equations. In general, systems of linear equations have high variability in tile occupancy because of this dense diagonal. This typically leads to poor buffer utilization with CST approaches due to a small number of tiles having high occupancy while the majority of tiles have low occupancy. 

Although some sparse linear solvers do involve multiple sparse operands~\cite{amg, schur}, most sparse linear solvers rely on sparse-dense tensor algebra. Thus, we also select tensors from other applications that rely more heavily on sparse-sparse tensor algebra such as graph and data analytics~\cite{msbfs,markov}. We focus on tensors that cannot fully fit inside the global buffer as tiling provides little benefit when all data fits on-chip.

\section{Evaluation}
\subsection{Comparison to ExTensor-N and ExTensor-P}

Fig.~\ref{fig:speedup} shows the speedup relative to ExTensor-N on all workloads for ExTensor-P and ExTensor-OB. ExTensor-OB has an average speedup of $52.7\times$  and $2.3\times$ over ExTensor-N and ExTensor-P, respectively, based on the impact of overbooking compared to prescient CST. Because ExTensor-P and ExTensor-OB construct tiles dependent on sparsity rather than with a fixed tile size, they are able to significantly improve on ExTensor-N in terms of both speed and efficiency. Because of this, we will primarily focus on the comparison between ExTensor-OB and ExTensor-P. We do not evaluate the preprocessing cost of ExTensor-P, but note that prescient preprocessing requires many iterations over the operand tensors to determine the optimal tile size and shape. 

Since Tailors enable tiles with occupancy greater than the available buffer capacity, the tiles used by ExTensor-OB are larger than those used by ExTensor-P. This leads to greater average buffer occupancy and improved data reuse per buffer fill, reducing expensive accesses to DRAM for tensors with more variation in sparsity. 

Because overbooking takes advantage of variability in tile occupancy, ExTensor-OB sees large speedups of $6.3\times$ and $5.7\times$ over ExTensor-P on tensors with very high variability such as \emph{roadNet-CA} and \emph{webbase-1M}, while workloads with uniformly distributed sparsity such as \emph{web-Google} and \emph{patents\_main} show similar speedup between ExTensor-P and ExTensor-OB compared to ExTensor-N. With less variability in the sparsity distribution, overbooking provides less benefit since allowing for overbooking does not significantly increase the tile size supported by the buffer. For these workloads, inaccuracy with Swiftiles' predictions can cause ExTensor-OB to perform worse than ExTensor-P due to inaccuracy in tile size estimation (\eg, \emph{email-Enron}, \emph{sx-askubuntu}). Workloads that fit almost entirely on chip such as \emph{sx-mathoverflow} and \emph{p2p-Gnutella31} also show very similar speedup between ExTensor-P and ExTensor-OB due to the reduced impact of tiling when most of the tensor already fits on chip.

Fig.~\ref{fig:energy} shows the energy consumption of ExTensor-P and ExTensor-OB relative to ExTensor-N. ExTensor-OB achieves a $22.5\times$ and $2.5\times$ reduction in energy compared to ExTensor-N and ExTensor-P, respectively. Overbooking is able to reduce energy even when unable to increase speed (\eg, \emph{email-Enron}) by allowing larger PE-level tiles and thus reducing accesses to the global buffer. Since ExTensor-OB is still limited by DRAM traffic, ExTensor-OB would see no speedup from these larger PE-level tiles. 

Dynamic reflexive tiling (DRT)~\cite{odemuyiwa_drt_2023}, which is concurrent with this work, proposed improving buffer utilization by constructing tiles dynamically based on sparsity. When compared to Tailors and Swiftiles, DRT requires more complex logic on-chip to facilitate dynamic tiling at runtime. To compare overbooking to DRT, we used the DRT simulator~\cite{odemuyiwa_drt_2023}, and found that ExTensor enhanced with DRT is $2.4\times$ faster than ExTensor-P. Then using our Sparseloop simulations~\cite{timeloop, sparseloop}, we found that ExTensor-OB is $2.3\times$ faster than ExTensor-P. Therefore, we extrapolate that ExTensor-OB is approximately the same speed as ExTensor with DRT, but with simpler hardware.

\subsection{Impact on Data Reuse}

\begin{figure}[t]%
    \centering
       \label{fig:dram-traffic}
\subfloat[]
{
\includegraphics[width=\columnwidth]{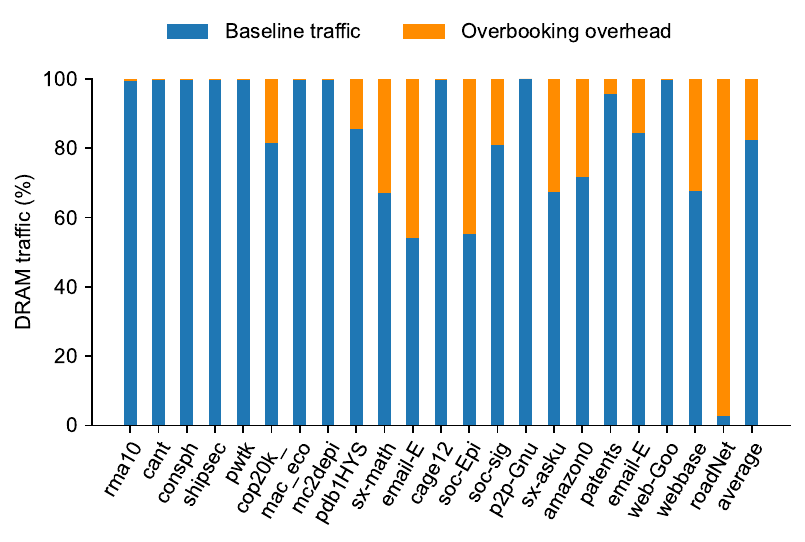}
\label{fig:dram-traffic:a}
}\\
\subfloat[]{
\includegraphics[width=\columnwidth]{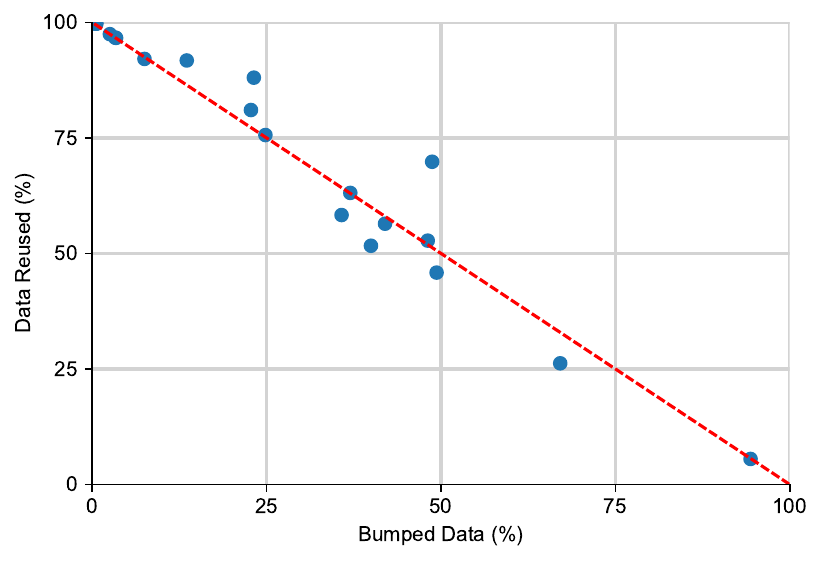}
   \label{fig:dram-traffic:b}
}
\vspace{-8pt}
    \caption{\textbf{Impact of overbooking on data reuse for different workloads. (a) Proportion of DRAM traffic used by streaming in \obsb~when 10\% of tiles overbook the buffer. The overhead in additional DRAM traffic of overbooking depends on the variation in the sparsity of each workload. (b) Percentage of data reused relative to the percentage of bumped data using Tailors when $y=10\%$. Each blue dot corresponds to a workload from SparseSuite. The strong correlation between data reuse and the bumped data (shown in red) indicates that Tailors is adaptable to many workloads instead of taking advantage of specific sparsity patterns for each workload. }}
\end{figure}

Overbooking affects data reuse in two ways: (1) in non-overbooked tiles, increasing the tile size leads to more reuse within a tile; however, (2) in overbooked tiles the portion of overbooked elements must be fetched from the parent buffer for every use and thus gets minimal reuse. When a tile overbooks the buffer, Tailors stream in the overbooked portion of the tile and do not exploit reuse on that overbooked portion. Thus, overbooking can be described both in terms of how many tiles are overbooked (\eg, $y=10\%$) and how much of each such tile is overbooked. Although Swiftiles targets a fixed percentage of overbooked tiles (\ie, how \emph{many} tiles are overbooked), the percentage of data that is bumped (\ie, how \emph{much} of the tile is overbooked) can vary between workloads depending on the sparsity distribution. Moreover, the degree of exploitable data reuse may vary based on specific sparsity patterns in the data.

Although ExTensor-OB's larger tiles increase average buffer occupancy and thus data reuse per buffer fill, support for overbooking results in some buffer fills with limited to no reuse. We use the percentage of DRAM traffic dedicated to streaming bumped data to study how the cost associated with overbooking varies across workloads.

Fig.~\ref{fig:dram-traffic:a} shows the DRAM traffic of streaming bumped data through the \obsb~relative to the baseline DRAM traffic assuming the same tiling and an infinitely large buffer that never overbooks. On average, overbooking of 10\% of tiles leads to 26\% overhead for streaming data because of the lost data reuse for bumped data in an overbooked buffer. This penalty is offset by the increased data reuse across other tiles due to the larger tile size enabled by overbooking.

For diagonally-dense coordinate-dependent tensors such as \emph{rma10}, \emph{cant}, and \emph{consph}, the traffic from streaming for overbooking is negligible as overbooking is unable to make much impact on tile size with $y=10\%$. Notably, although these tensors have high variability in tile occupancy, the tile occupancy distribution is very deterministic: the region along the diagonal has many nonzeros, while the region away from the diagonal has very few nonzeros.

Some tensors such as \emph{roadNet-CA} see baseline DRAM traffic get dominated by accesses to bumped data. This is because \emph{roadNet-CA} has a highly asymmetric tile occupancy distribution, that is, that there are very few tiles that each have very high occupancy and many tiles with very low occupancy.

Another way to show the impact of overbooking on data reuse is by comparing the percentage of data that is treated as bumped data to the percentage of data that is reused (Fig.~\ref{fig:dram-traffic:b}). If all tiles fit without overbooking, the percentage of data reused would be $100\%$ since any output could be computed from values already held in the buffer. As fewer tiles fit and more data in each tile is overbooked, the percentage of data reused would approach $0\%$ due to the smaller likelihood of data accesses matching data held in the buffer.

The comparison shown in Fig.~\ref{fig:dram-traffic:b} isolates the impact of how much each tile is overbooked and helps understand how sparsity variation within a tile impacts overbooking with Tailors. Specifically, since Tailors keeps a fixed portion of tile data resident in the buffer (\ie, the first elements that fit), sparsity patterns in tensors may cause the data in the buffer to be accessed rarely or never accessed. This can occur when the coordinates of nonzeros in the one operand intersect only  the coordinates of \emph{overbooked nonzeros} in the other operand. Tailors introduce no mechanism for replacing different data if the portion of the tile held in the buffer sees limited reuse.

We observe that data reuse and the percentage of bumped data are strongly correlated. The strong correlation between data reuse and the percentage of bumped data shows that Tailors' efficacy depends primarily on the percentage of bumped data as expected from a scan access pattern rather than on specific sparsity patterns from each workload. Although Tailors is not able to fully exploit data reuse for some sparsity patterns, the likelihood of sparsity patterns that harm data reuse in Tailors (\ie, by accessing data not in the buffer more often than data in the buffer) is no greater than the likelihood of sparsity patterns that benefit data reuse. The use of different replacement policies, specifically those that manage data replacement with greater flexibility for what data is kept in the buffer, may improve data reuse and is an interesting direction for future work. For example, instead of using the end of the buffer as the FIFO-managed region, the region used for replacement could be selected on a per-workload or per-tile basis or could adopt a different replacement policy (\eg, LIFO) with corresponding changes to data orchestration.

\subsection{Impact of Swiftiles Parameters}

Swiftiles introduces a number of parameters that can be tuned for improved prediction accuracy as well as improved performance. In this section, we study the behaviour of Swiftiles for the different parameters. 
\label{sec:swiftiles-parameters}

\begin{figure}[t]
\centering
\includegraphics[width=\columnwidth]{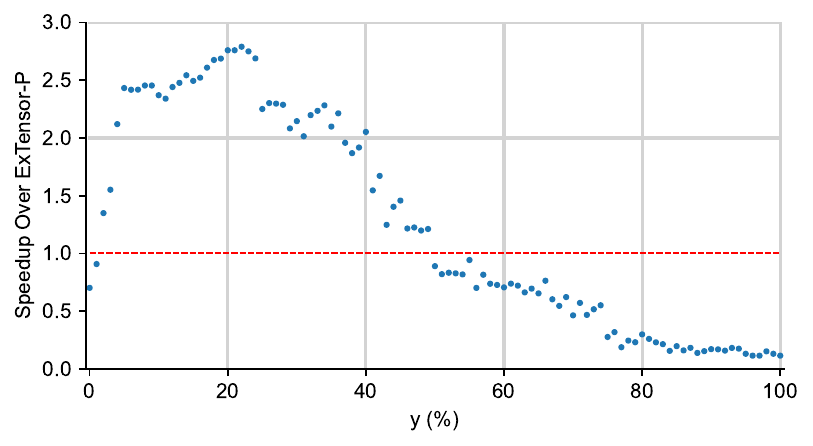}
\caption{\textbf{Speedup of ExTensor-OB over ExTensor-P using Swiftiles with different overbooking probabilities $y$. The speedup is averaged across all workloads. We show ExTensor-P in red for comparison. The choice of $y=10\%$ falls in a region that is relatively insensitive to changes in overbooking rate.}}
\label{fig:ysweep}
\end{figure}

\begin{figure}[t]
\centering
\includegraphics[width=\columnwidth]{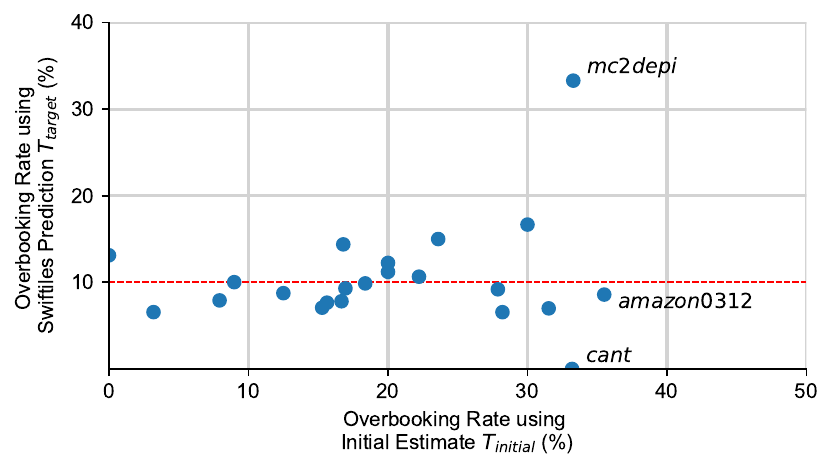}
\caption{\textbf{Comparison of the overbooking rate between tiling the tensor using the initial estimate and tiling with the Swiftiles predicted tile size when the target $y=10\%$ (shown in red) is used and all tiles are sampled. Each blue dot corresponds to a workload from SparseSuite. }}
\label{fig:scaling}
\end{figure}

\begin{figure}[t]
\centering
\includegraphics[width=\columnwidth]{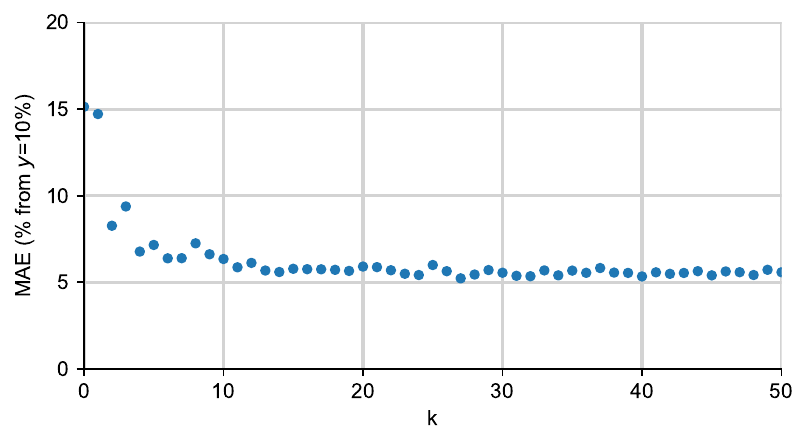}
\caption{\textbf{MAE of Swiftiles predictions as the number of samples increases and $y=10\%$. With $k=0$, no sampling occurs and Swiftiles uses the initial estimate. Based on Swiftiles, the total number of tiles sampled is equal to $10 \times k$. As the number of samples increases, Swiftiles predictions converge to a certain degree of error. Swiftiles does not converge to 0 MAE because Swiftiles only samples for one tile size. }}
\label{fig:k}
\end{figure}

\begin{figure*}[t]
\centering
\includegraphics[width=\textwidth]{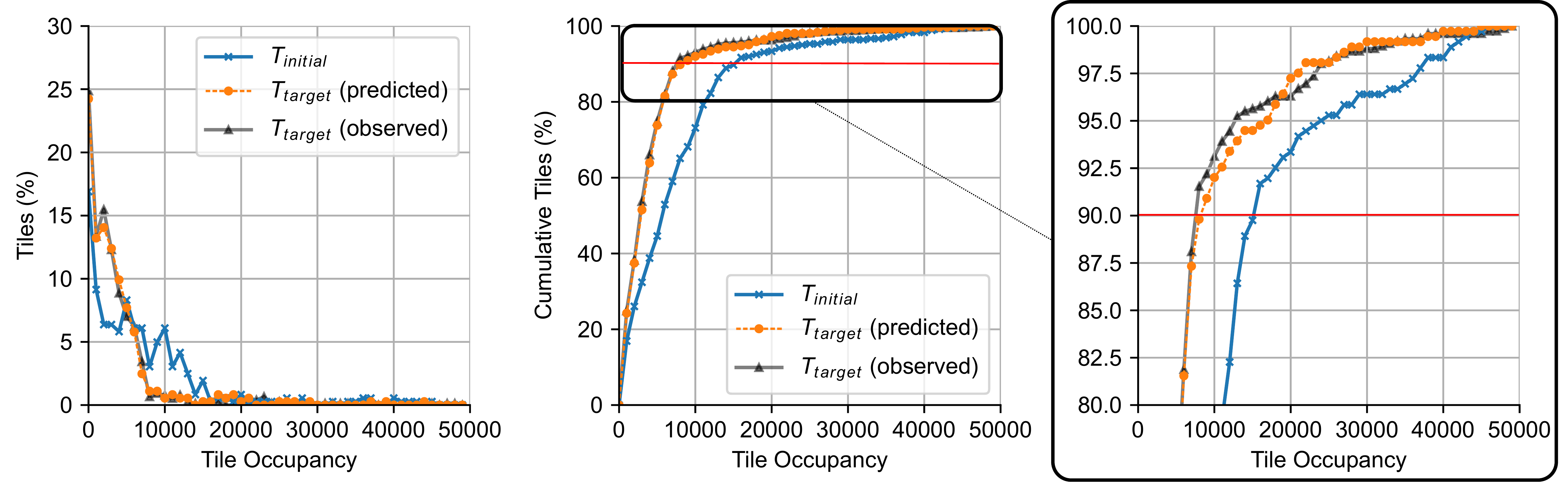}

\caption{\textbf{Tile occupancy distributions for Swiftiles applied on the workload \emph{amazon0312} when targetting a buffer size of 8K nonzeros and $y=10\%$. The distribution made when tiling with the initial estimate is shown as $T_{initial}$, the scaled distribution created by Swiftiles is shown as $T_{target}$(predicted), and the actual distribution observed when tiling with the target tile size is shown as $T_{target}\text{(observed)}$. (a) The probability density function of tile occupancies. (b) The cumulative distribution function of tile occupancies. (c) The cumulative distribution function, specifically when $80\%$ to $100\%$ of tiles fit in the buffer. The $y=10\%$ point (90\% of tiles fit) is shown in red. }}
\label{fig:amazon0312}
\phantomsubcaption
\label{fig:amazon0312:a}
\phantomsubcaption
\label{fig:amazon0312:b}
\phantomsubcaption
\label{fig:amazon0312:c}
\end{figure*}

\textbf{Impact of $y$:} The selection of $y$ makes a key assumption for how much overbooking is desirable when tiling. To evaluate the efficacy of our choice of $y$, we compare the speedup of ExTensor-OB over ExTensor-P with different values of $y$ in Fig.~\ref{fig:ysweep}. 

At $y=0\%$ when Swiftiles predicts no tile as overbooked, ExTensor-OB is approximately 25\% slower than ExTensor-P due to inaccuracy in tile size estimates from Swiftiles. As $y$ increases up to $22\%$, ExTensor-OB selects progressively larger tile sizes and gets faster due to increasing buffer utilization.  As $y$ increases past 22\%, ExTensor-OB begins to select tile sizes for which the overbooking overhead exceeds the benefit from improved buffer utilization and thus reduces performance. At $y=100\%$, Swiftiles predicts every tile as overbooked and ExTensor-OB performs significantly worse than ExTensor-P as it pays the data reuse penalty for overbooking every tile. We select $y=10\%$, which falls in a region that is relatively insensitive to variations in $y$. 

To give an idea of the impact of using a fixed $y$ across all workloads, we further compare to an idealized version of ExTensor-OB that selects the best $y$ for each workload. We find that this idealized version of ExTensor-OB is $4.8\times$ faster than ExTensor-P and $2.1\times$ faster than ExTensor-OB with $y=10\%$. ExTensor-OB loses half of its potential performance due to the static selection of $y$ across all workloads; however, similar to ExTensor-P, selecting the best $y$ for each workload would incur a significant preprocessing overhead due to having to check the occupancy of each tile at all tile sizes as well as searching for $y$.

\textbf{Impact of scaling:} Swiftiles relies on the assumption that tile occupancy distributions do not change for small variations in tile size. As shown in Fig.~\ref{fig:swiftiles:b}, the sampled distribution generated from the initial estimate is used to identify the tile occupancy that 10\% of sampled tiles exceed. After scaling, the expectation is that the overbooking rate will average the target $y=10\%$. To evaluate the performance of the tile estimator, we compare the error between the average overbooking rate across different workloads and the target as well as the variation of overbooking rate to the target.

Fig.~\ref{fig:scaling} compares the overbooking rate for different workloads using the initial estimate $T_{initial}$ and the final predicted tile size $T_{target}$ when $y=10\%$. Tiling with the initial estimate leads to an average overbooking rate of $19.9\%$ and a mean average error (MAE) of $15.6\%$ across the workloads we study in SparseSuite. Notably, the average overbooking rate with the initial estimate is significantly different from the target $y=10\%$ as the initial estimate makes no effort to approximate the tile occupancy for a given $y$. After scaling with Swiftiles, the average overbooking rate is $10.6\%$ with an MAE of $5.8\%$, matching $y$ on average and significantly reducing error. Due to variations in sparsity characteristics, different workloads behave differently when scaled. In particular, the tile occupancy distribution of workloads such as \emph{cant} and \emph{mc2depi} are poorly approximated by the initial estimate and do not scale linearly with tile size, leading them to deviate from the $y=10\%$ target.

\textbf{Impact of $k$:} When constructing the sample distribution, there exists a tradeoff between sample distribution accuracy and the cost of collecting more samples. In order to evaluate the ideal number of samples Swiftiles should collect to construct the tile occupancy distribution, we compare the MAE of Swiftiles' predictions using different $k$ averaged across all workloads. 

Fig.~\ref{fig:k} shows the MAE of Swiftiles predictions as the number of positive samples collected varies from no samples to fully sampling all tiles. Although error decreases as the number of samples increases, there are diminishing returns to increasing the number of samples. With $k=10$, MAE is 5.8\%, compared to 5.5\% when all tiles are sampled. The gap that remains between the fully-sampled Swiftiles estimate and the actual target is caused by the one-shot process of Swiftiles: Swiftiles only checks one tile size (the initial estimate) before making a prediction to maintain the low cost of preprocessing. 

An example of the Swiftiles process is shown in Fig.~\ref{fig:amazon0312}, which compares the tile occupancy distributions gathered by Swiftiles to the observed tile occupancy distribution when the tensor \emph{amazon0312} is tiled with tile size $T_{target}$. Fig~\ref{fig:amazon0312:a} shows the scaling process from Swiftiles: given the initial estimate $T_{initial}$ and a number of samples, Swiftiles scales the tile occupancy distribution so that 90\% of tiles contain less than 8K nonzeros. Fig.~\ref{fig:amazon0312:b} shows the cumulative distribution function of the given distributions to better visualize the impact of scaling on the overbooking rate. Despite the relative inaccuracy of $T_{initial}$, scaling helps the distribution $T_{target}$~(predicted) align with $T_{target}~\text{(observed)}$.

\section{Related Work}
\subsection{Concept of Overbooking}
Overbooking is a widely used approach in various industries for cost savings and improvements in efficiency when faced with limited resources~\cite{weatherford1992taxonomy}. For instance, airlines~\cite{rothstein1971airline}, deliberately overbook planes to minimize the loss incurred by cancellations and `no-shows', while clinics overbook to increase patient access~\cite{laganga2007clinic}. While algorithms used to determine the amount of overbooking for these applications can be quite complex~\cite{rothstein1974hotel, zacharias2014appointment}, our Swiftiles is a relatively simple approach. In addition, our Tailors ensure that all tiles reach their destination (\ie, no denied service), avoiding disastrous overbooking scenarios that we know all too well (\eg,~\cite{presale}). 

Overbooking has also been explored in other aspects of computing including overbooking CPU and networking resources in the data center to improve utilization~\cite{urgaonkar2002resource, tomas2013improving}. In this work, we overbook storage resources in an accelerator to improve buffer utilization.

\subsection{Tiling Strategies and Storage Idioms}
To the best of our knowledge, tiling strategies for sparse tensor algebra workloads have not been widely studied. ExTensor~\cite{hegde_extensor_2019} proposed to perform CST across the entire tensor. Dynamic Reflexive Tiling (DRT)~\cite{odemuyiwa_drt_2023}, which is concurrent with our overbooking work, performs coordinate-based position-space tiling. However, DRT introduces complicated and expensive tile construction control to search for tiles in position space and has significant overhead. 
To the best of our knowledge, no prior work has explored coordinate-space tiling where tiles may not fit within a given buffer.

There also exist various storage idioms and buffering strategies for domain-specific accelerator designs~\cite{jenga, desc, pdae, stash, acceleratorstore, leapscratchpads, coram, stream-dataflow, pellauer_buffet_2019}. However, none of them allow data allocation to a buffer to exceed the buffer capacity to efficiently support overbooking.

\subsection{Existing Sparse Tensor Accelerators}
There is ample prior work designing accelerators for efficiently processing various sparse tensor algebra workloads~\cite{hegde_extensor_2019, outerspace, matraptor, sparch, sigma, eyeriss, sparse-reram-engine, vector-sparse-tensor-core, s2ta, eyeriss-v2, procrustes}. However, these works focus on enabling flexible sparsity support by designing novel sparse dataflows or performing software-hardware co-design with novel sparsity patterns. Such proposals are often complementary to tiling strategy choices, which is the focus of our work and can therefore be integrated with prior work.

The GAMMA accelerator~\cite{GAMMA-accelerator} has some similarities to this work in terms of managing data overflow of the buffer to achieve similar benefits to overbooking, but differs from Tailors in three key aspects: (1) Tailors uses explicit data orchestration, while GAMMA uses implicit data orchestration; (2) Tailors supports streaming of tiles of both operands, while GAMMA only streams row data for the non-stationary operand; and (3) Tailors performs coordinate-space tiling of both operands, while GAMMA only performs selective coordinate-space tiling of very high-occupancy rows of the stationary operand. 

\section{Conclusion}
Tiling is key to improving data reuse and thus reducing memory traffic for sparse tensor algebra applications. This paper addresses the importance of balancing the tiling strategy's adaptability and efficiency by proposing a speculative tiling strategy, Swiftiles, that achieves high buffer utilization by constructing tiles that occasionally overbook the available buffer capacity. By statistically estimating tensor sparsity characteristics, Swiftiles introduces minimal preprocessing overhead. In conjunction, we integrate a low-overhead hardware recovery mechanism, Tailors, into the existing memory hierarchy to ensure correctness for tiles that overbook the buffers. Across representative workloads, we demonstrate that allowing overbooked tiles can introduce a $2.3\times$ speedup and a $2.5\times$ reduction in energy compared to existing accelerators. We think it possible that the overbooking paradigm can be extended beyond buffers in sparse tensor accelerators, including overbooking of data conversion in resistive memories and overbooking of compute elements in machine learning accelerators. We hope that this work inspires research on the use of overbooking in these other spaces.

\begin{acks}
We would like to thank the anonymous reviewers for their constructive feedback. This research was funded in part by the MIT AI Hardware Program. We would like to thank Nandeeka Nayak and Toluwanimi Odemuyiwa for their help in enabling us to better validate/extend ExTensor and DRT, respectively.


\end{acks}

\balance
\bibliographystyle{ACM-Reference-Format}
\bibliography{refs}

\end{document}